\documentclass[12pt,preprint]{aastex}

\newcommand{\rplotone}[2]{\epsscale{#1}\plotone{#2}\epsscale{1.0}\\}

\newcommand{\uchii}{UC\ion{H}{2}}

\shorttitle{Clumpy Protostellar Models}
\shortauthors{Whitney IV}

\begin{document}

\title{3-D Models of Embedded High-Mass Stars:  Effects of a Clumpy Circumstellar Medium}

\author{R. Indebetouw\altaffilmark{1,4},
B. A. Whitney\altaffilmark{2},
K. E. Johnson\altaffilmark{1,5},
K. Wood\altaffilmark{3}}

\altaffiltext{1}{University of Virginia, Astronomy Dept., P.O. Box 3818, Charlottesville, VA, 22903-0818}
\altaffiltext{2}{Space Science Institute, 4750 Walnut St. Suite 205,
Boulder, CO 80301}
\altaffiltext{3}{School of Physics and Astronomy, University of St. Andrews, North Haugh, St Andrews, Fife, Scotland, KY16 9SS}
\altaffiltext{4}{Spitzer Fellow}
\altaffiltext{5}{Hubble Fellow}

\begin{abstract}

We use 3-D radiative transfer models to show the effects of clumpy
circumstellar material on the observed infrared colors of high mass
stars embedded in molecular clouds.  
We highlight differences between 3-D clumpy and 1-D smooth models
which can affect the interpretation of data.
We discuss several important properties of the emergent spectral
energy distribution (SED): More near-infrared light (scattered and
direct from the central source) can escape than in smooth 1-D models.
The near- and mid-infrared SED of the same object can vary
significantly with viewing angle, depending on the clump geometry
along the sightline.  Even the wavelength-integrated flux can vary
with angle by more than a factor of two. Objects with the same {\it
average} circumstellar dust distribution can have very different near-
and mid-IR SEDs depending on the clump geometry and the proximity of
the most massive clump to the central source.

Although clumpiness can cause similar objects to have very different
SEDs, there are some observable trends.  Near- and mid-infrared colors
are sensitive to the weighted average distance of clumps from the
central source and to the magnitude of clumpy density variations
(smooth-to-clumpy ratio).  Far-infrared emission remains a robust
measure of the total dust mass.  We present simulated SEDs, colors,
and images for 2MASS and {\it Spitzer} filters.  We compare to
observations of some \uchii\ regions and find that 3-D clumpy models
fit better than smooth models.  In particular, clumpy models with
fractal dimensions in the range 2.3-2.8, smooth to clumpy ratios of
$\lesssim$50\%, and density distributions with shallow average radial
density profiles fit the SEDs best ($\left<\rho\right>\propto
r^{\alpha}, -1.0<\alpha<0.0$).

\end{abstract}



\section{Introduction}

The important question of how stars form is frequently investigated
with infrared imaging, for both astrophysical and practical reasons.
Young stellar objects (YSOs) are usually still deeply embedded in their natal
molecular material.  The distribution of this gas and dust (e.g. in an
accretion disk) can reveal a great deal about the formation process,
and this hot dust close to the star emits primarily in the near
and mid-infrared (for this paper, we define near-IR
$\equiv$1$<\lambda<$3$\mu$m and mid-IR $\equiv$3$<\lambda<$30$\mu$m).
Although the bulk of the dust mass is often at cooler temperatures,
with SEDs peaking in the far-infrared ($\sim$100$\mu$m), probing the
warmer dust is both advantageous and potentially dangerous because the
NIR/MIR SED is more sensitive to the dust distribution and geometric
effects.
The infrared is observationally advantageous because at shorter
wavelengths, YSOs are completely invisible, and at longer wavelengths,
detector technology is less mature and the spatial resolution of
(typically diffraction-limited) observatories begins to suffer.  Many
current and recent observatories are infrared-optimized, such as ISO
and ground-based telescopes with adaptive optics. Particularly
exciting is the successful operation of the {\it Spitzer Space
Telescope}, which is producing a wealth of MIR data on star formation
due to its unprecedented mapping efficiency
\citep{allen04,megeath04,reach04,barbsup}.  The IRAC instrument
\citep{fazio04} provides simultaneous $\lesssim 2\arcsec$ imaging at
3.6, 4.5, 5.8, and 8.0 $\mu$m, and MIPS \citep{rieke04} provides
6\arcsec images at 24$\mu$m, and longer wavelengths at lower
resolution.  It is critical that we understand how circumstellar dust
geometry can affect our interpretation of all of this NIR/MIR data.

In a previous set of papers \citep{pI,pII,pIII} we showed how 2-D
geometries (disks, rotationally flattened envelopes, bipolar cavities)
affect the near-IR and mid-IR colors of Young Stellar Objects (YSOs)
of various stellar masses and evolutionary states.  Those models show
that there is not always a direct correspondence of the 1-10 $\mu$m
colors or spectral index with evolutionary state, due to several
effects such as scattered light, inclination, aperture size, and
stellar temperature.  There is a trend for younger sources to have
redder colors, but there is substantial overlap and extreme exceptions
to the trends.  Separation of these populations in a color-color
diagram generally requires supplemental longer ($\lambda>20\mu$m)
wavelength observations.  If only $1-10 \mu$m SEDs are available, for
example, in the Galactic Legacy Infrared Midplane Extraordinaire
\citep[GLIMPSE, a {\it Spitzer} project;][]{pasp}, model fits to the
spectral energy distributions (SEDs) provide a more promising avenue
than color-color diagrams for interpreting the SEDs since they use the
most information simultaneously (Robitaille et al. 2005, in prep.).

This paper focuses on UltraCompact \ion{H}{2} regions, or the later
stages of high mass (O and early B) star formation; although we have
not yet explored the issue, we expect that the clumpiness of envelopes
is important also in younger high-mass objects, and perhaps in
somewhat lower-mass protostars as well.  It is believed that the
central sources of UltraCompact\ion{H}{2} regions
\citep[{\uchii}s;][]{edsup} have probably reached the main sequence,
possibly halted accretion and dispersed infalling material, but remain
deeply embedded in their natal molecular cloud.  These stars are so
luminous that they heat up large volumes ($\sim 3$ pc radius) and
masses of dust and gas, which reprocess the stellar radiation towards
longer wavelengths.  Observations of high mass YSOs and \uchii\
regions are usually modeled using 1-D radiative transfer codes
\citep{wolfire,faison,hatchell,vandertak,beuther,mueller}.  These are
adequate for modeling long wavelength observations ($> 100 \mu$m), but
at shorter wavelengths, the inhomogeneous nature of the surrounding
interstellar medium \citep[e.g.][]{clumpyISM} likely has a large
effect on the SED.  This is demonstrated in the attempts by
\citet{faison} (hereafter F98) to fit the mid-IR spectroscopy of
several \uchii\ sources with 1-D smooth models.  For a given envelope
mass set by the long wavelength flux, the models consistently produced
too deep a 10 $\mu$m silicate feature and too little short wavelength
flux, even after decreasing the silicate abundances by a factor of
two.  Changing the radial density profile is insufficient to bring the
models into agreement with the data.  F98 use constant-density spheres
with large inner holes and obtain a poor match.  Some of the other
models cited above use negative power-law density gradients to fit
spatially resolved intensity profiles; such 1-D models still have
difficulties fitting \uchii\ SEDs over a wide wavelength range
(\S\ref{gradient}).

In \S\ref{models} we describe our models.  Three-dimensional
nonhomogeneous models differ significantly from 1-D smooth models with
the same amount of circumstellar dust.
Some authors have computed SEDs in two-phase clumpy distributions
\citep{misselt,wolf}. In an attempt to calculate as realistic models
as possible, in this paper we calculate SEDs in hierarchically clumped
density structures, following a procedure for calculating
fractal density described by \citet{elmegreen97} and \citet{refneb}
and using a volume fractal dimension $D$ similar to that calculated
for the interstellar medium \citep[$D=2.3-2.6$;][]{clumpyISM}.
In \S\ref{results} we describe the properties of the 3-D models in
terms of synthetic observations: high-resolution images, spectral
energy distributions, and photometric fluxes analyzed using
color-color and color-magnitude diagrams.  We discuss how the
properties vary with viewing angle (\S\ref{viewing}), and show in
detail how the SEDs of 1-D models cannot reproduce the characteristics
of 3-D models (\S\ref{ave}).  In \S\ref{trends} we show that despite
the variations, there are some trends that allow observations to probe
physical conditions. In \S\ref{wccolors} we demonstrate our findings
on real data, and give conclusions in \S\ref{conclusions}.


\section{Model Construction}
\label{models}

We use a Monte Carlo Radiative transfer code similar to that described
in \citet{pI} (Paper I), but using a 3-D grid and incorporating a
hierarchically clumped density distribution.  We use the radiative
equilibrium method developed by \citet{bjork01}.  The models solve for
the 3-D temperature distribution, conserve flux absolutely, and
accurately compute scattering and polarization using arbitrary
scattering phase functions.  The output radiation field is binned into
200 angular directions (10 polar intervals and 20 azimuth intervals),
i.e.  each model produces 200 SEDs corresponding to different viewing
angles.
We do not currently include emission from polycyclic aromatic
hydrocarbons (PAHs) or transiently heated grains; these likely affect
fluxes in the IRAC wavelength range (3--8$\mu$m).  The only source of
calculation error in our models is photon counting statistics --
running more photons produces higher signal-to-noise spectra.  The
models produced for figures in this paper took 12--24 hours each to
run (on a 2~GHz PC and code compiled with g77).  Draft quality
results sufficient to explore trends take $\lesssim$2 hours.

We specify circumstellar dust distributions with a clumpy and smooth
component, with the relative contribution ranging from zero (very
clumpy with evacuated holes) to one (constant density).  The clump
distribution is specified using the algorithm of \citet{elmegreen97}
as implemented in \citet{refneb}: $N$ mass particles are randomly
placed in a simulation grid of size $L$.  $N$ more particles are
randomly placed, each within a distance $L/(2\Delta)$ of one of the
particles in the previous round ($\Delta$ is a parameter which
determines the approximate fractal dimension of the resulting
distribution).  This process is repeated twice more, so that $N^4$
particles are placed.  The mass distribution is not a true fractal
(self-similar on all scales), but merely hierarchically clumped
(self-similar over about an order of magnitude in scale).
Nevertheless, the quantity $D=\log(N)/\log(\Delta)$ is essentially the
fractal dimension.  The precise value of $D$ does not dramatically
affect any of our results -- this will be demonstrated in
\S\ref{fracdimsec}, after discussing more universal properties of the
models.  

Different grain models could be used in different parts of the dust
cocoon (e.g. larger grains in dense clumps), as in the young stellar
objects modeled in \citet{pII}.  For simplicity, we use the same grain
model throughout the envelope; as described in Paper~I, the grain
distribution was fit to the extinction curve typical of molecular
clouds with $R_V$=4.3 using a maximum entropy method similar to
\citet{kmh}.  We use spherical grains with dielectric functions for
silicate and graphite from \citet{laor}.  The dust sublimation
temperature is chosen to be 1600~K.

For simplicity, all of the clumpy models shown in this paper are for a
massive star, with the same amount of circumstellar dust, and the same
photospheric spectrum of the central source (the appropriate Kurucz
model).  The results presented are general to all massive young stars,
but we wish to isolate the effects of differing clumpy circumstellar
distribution from the effects of different stellar masses and
temperatures.  The latter effect is discussed in \citep{pII} for 2-D
circumstellar distributions, and further discussion of clumpy dust in
the context of embedded protoclusters is presented in Johnson et al
2005 (submitted).

\begin{figure}[h]
\rplotone{1.}{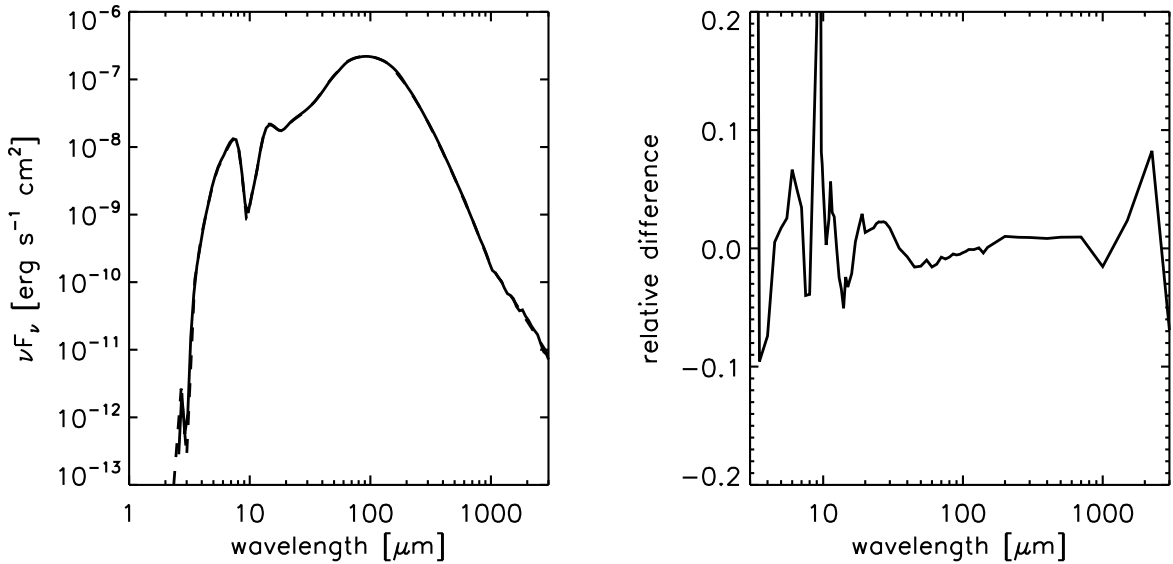}
\caption{\small\label{dusty} The first panel shows a 1-D model (spherically
symmetric) calculated with our code (solid line) and DUSTY (dashed
line).  The source is a main sequence 06-07 star ($T_\star$=41000~K)
in a 50000~$M_\sun$ dust cocoon extending from 0.0001 to 2.5~pc.  The
optical depth of the cocoon is $\tau_V$=117.5, and the temperature on
the inner radius is 700~K.  The second panel shows the fractional
difference between our model and one calculated with DUSTY.
Variations of a few percent are mostly due to noise in our spectrum.
}
\end{figure}

In order to provide a basis for comparison, we ran numerous models
with the smooth-to-clumpy density ratio equal to one,
i.e. constant-density spheres.  In particular, we compare our results
with the popular DUSTY code \citep{dusty} in Figure~\ref{dusty}.
Across a range of parameters, the SEDs are generally indistinguishable
between the two codes, and quantitatively agree to a few percent.  Our
code properly calculates nonisotropic scattering (Paper~1), whereas
DUSTY uses an isotropic scattering function.  For models with low
($\tau_V \lesssim 30$) optical depth, this can increase the
differences between the two codes to 10-15\%, but if we run our code
with isotropic scattering the results are then indistinguishable.  We
will demonstrate that differences between 1-D and 3-D models are far
greater than any of these subtle (few percent) effects.

\section{Results of a Canonical Clumpy Model}

\subsection{Density and Temperature Distributions}

\begin{figure}
\rplotone{0.7}{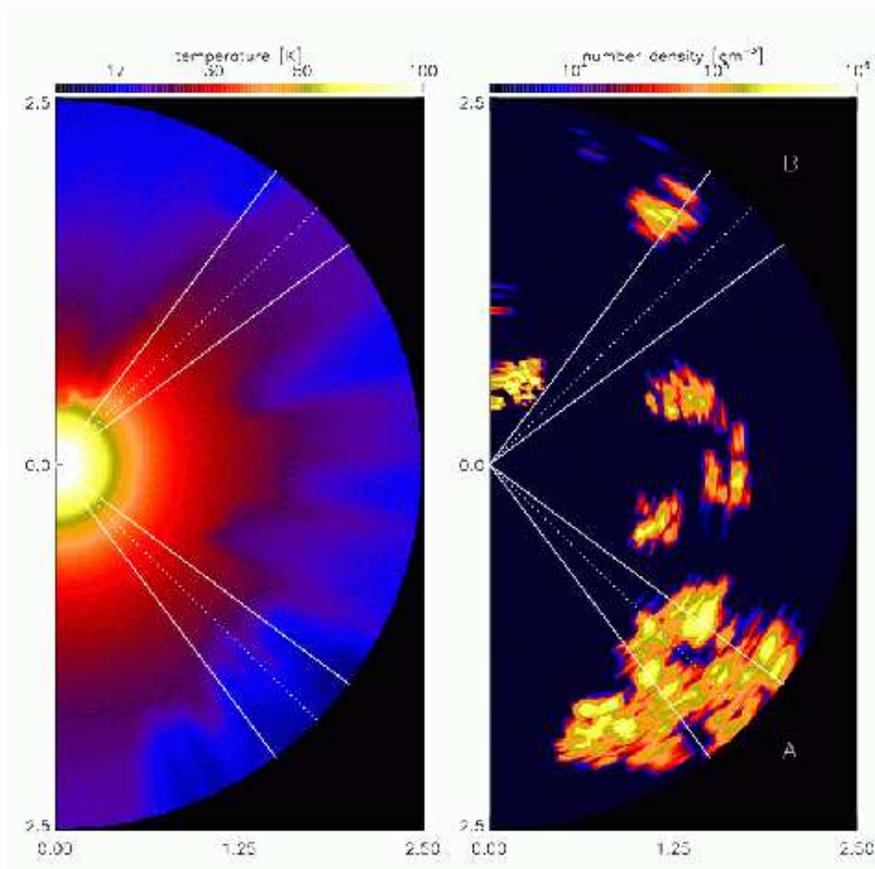}
\caption{\small\label{tempdens}Density and temperature distributions in a
slice of a typical model.  Both quantities are shown on a logarithmic
scale, with the ranges shown by the color bars. The spatial scale is
in parsecs.  Two particular sightlines discussed in the text (``A''
and ''B'', \S\ref{viewing}) are shown as dotted lines (although we
will discuss a sightline as if it were a single viewing angle,
radiation leaving the simulation is binned over the finite range of
angles indicated by the solid lines).}
\end{figure}

Figure \ref{tempdens} shows an azimuthal slice through the density
grid of our canonical model, a main sequence 06-07 star
($T_\star$=41000~K, L$_\star$=2.5$\times$10$^5$L$_\sun$) in a dust
cocoon extending from 0.0001~pc to 2.5~pc, with a smooth-to-clumpy
ratio of 0.1 and fractal dimension 2.6.  The cocoon is representative
of the inferred properties of massive YSOs: a thousand solar masses of
dust in about a 1/3 parsec region.  (At typical Galactic distances of
5kpc, 1/3pc=13\arcsec and 2.5pc=100\arcsec.)  As massive YSOs can heat
a large part of their natal clouds, we chose to include the larger
envelope out to 2.5pc and 50000~$M_\sun$; this does not qualitatively
affect our conclusions.  The visual extinction averaged over all
sightlines is $A_V$=131, but varies between 13 to 401 magnitudes
depending on viewing angle.  The voids at small radii allow stellar
radiation to propagate further, heating the inner faces of clumps.
This can result in a greater mass of warm dust than if all of the
stellar optical and UV radiation was intercepted near the star; in
other words, the $\tau_V=1$ surface is irregular and will enclose a
different amount of mass than in a spherically symmetric model.  This
is clearly visible in the temperature map, which shows cool dust in
the shadows of clumps, and heated inner surfaces.

\begin{figure}
\rplotone{0.5}{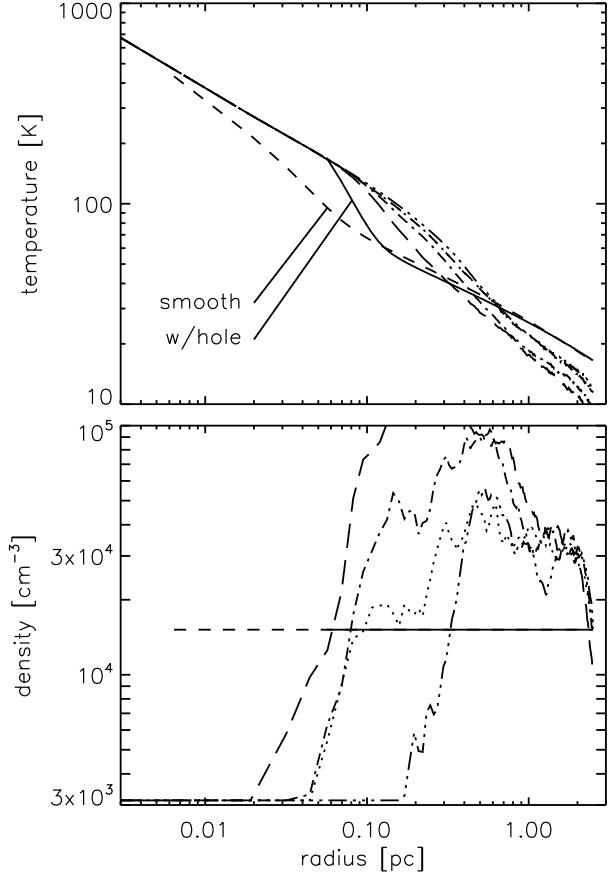}
\caption{\small\label{radtemp} Comparison of spherically-averaged radial
temperature and density profiles for smooth and clumpy models.  All
models shown here have the same amount of circumstellar dust within
the same 2.5~pc radius.  Two smooth models (solid and dashed lines)
and 4 clumpy models (long dashed, dot-dashed, dotted, and
triple-dot-dashed lines) are shown; the density and temperature
profiles for the clumpy models are averages over all angles. The upper
plot shows the temperature profiles, and the lower the corresponding
density profiles. The smooth model which extends all the way to the
dust destruction radius at constant density (dashed line) has the
shallowest gradient, because it has the highest density in the inner
regions of the four clumpy models.  The temperature of a smooth model
with a completely evacuated inner hole (solid line) also falls off
faster with radius than any clumpy model, because the inner faces of
clumps at varying distances are illuminated and heated, whereas all
stellar radiation of a given wavelength is intercepted at the same
radius in a smooth model.}
\end{figure}

The radial temperature profile is sometimes used as a diagnostic when
young stellar objects are modeled.  As one would expect, the radial
temperature profiles for clumpy models vary greatly depending on the
placement of the clumps of dust.  Even the radial temperature profiles
averaged over all angles show significant variation with the
particular clump geometry, and differ from the profile of a smooth
(1-D) model with the same amount of circumstellar mass.
Figure~\ref{radtemp} illustrates this comparison with smooth and
clumpy models that have the same amount of circumstellar mass.  The
smooth model's temperature gradient is shallower than the clumpy
models in the inner regions, because the clumpy models tend to have
lower average density there.  However, a smooth model with a
completely evacuated hole out to the distance of the nearest dense
clump in the clumpy models also shows a faster temperature dropoff
than the clumpy models.  This is the effect mentioned above, that the
$\tau$=1 surface for intercepting stellar radiation is irregular, and
clump surfaces can be heated at larger distances from the star than in
a spherically symmetric model with the most similar radial dust
distribution.  For the dust used here, temperature is expected to drop
off as $r^{-0.33}$ or $r^{-0.4}$ in the optically thin and thick
smooth cases, respectively \citep[more discussion in][]{pI}.  The
different slopes for different clumpy models likely reflect
transitions between optically thin interclump material and optically
thick clumps.

\subsection{SEDs}

\begin{figure}
\rplotone{0.65}{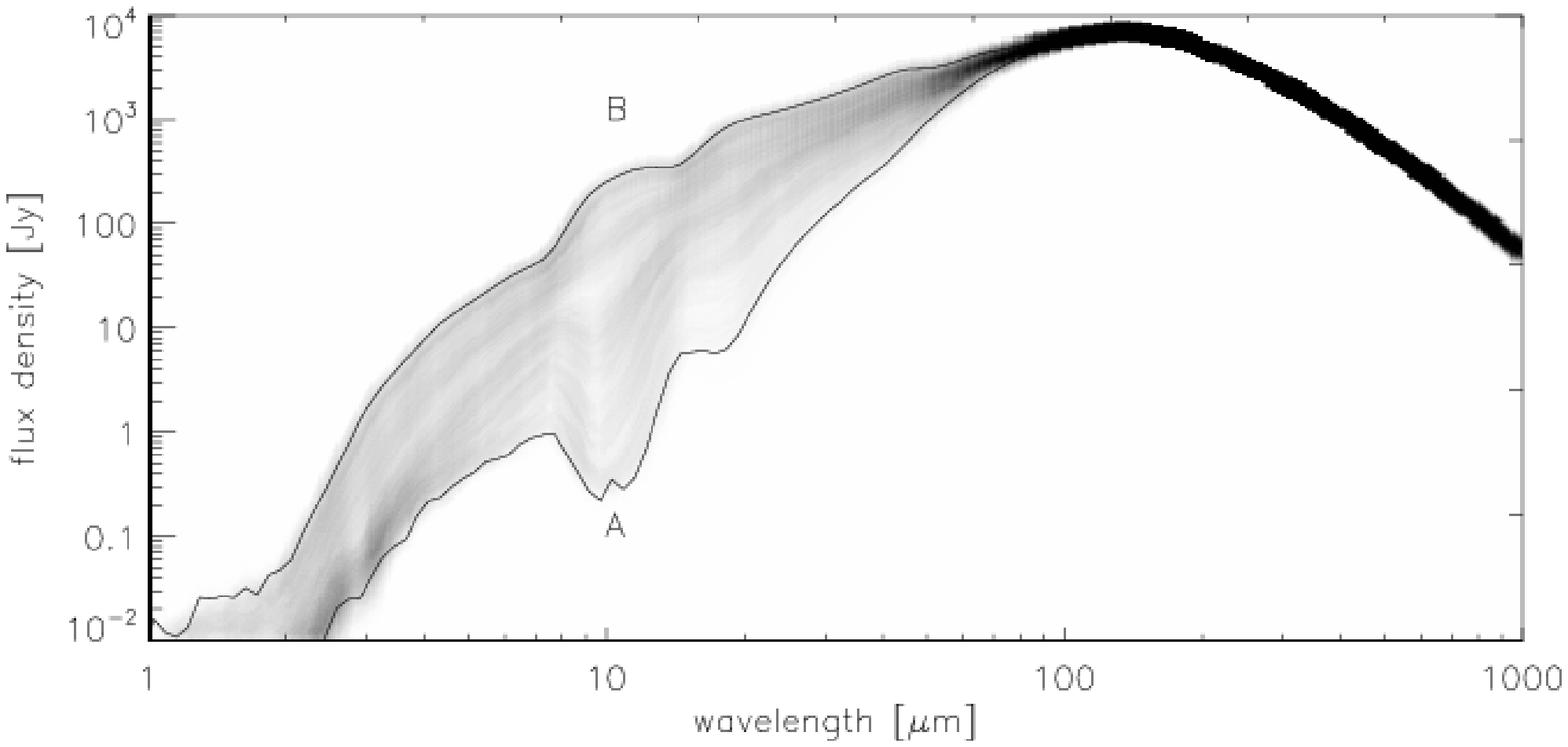}
\rplotone{0.65}{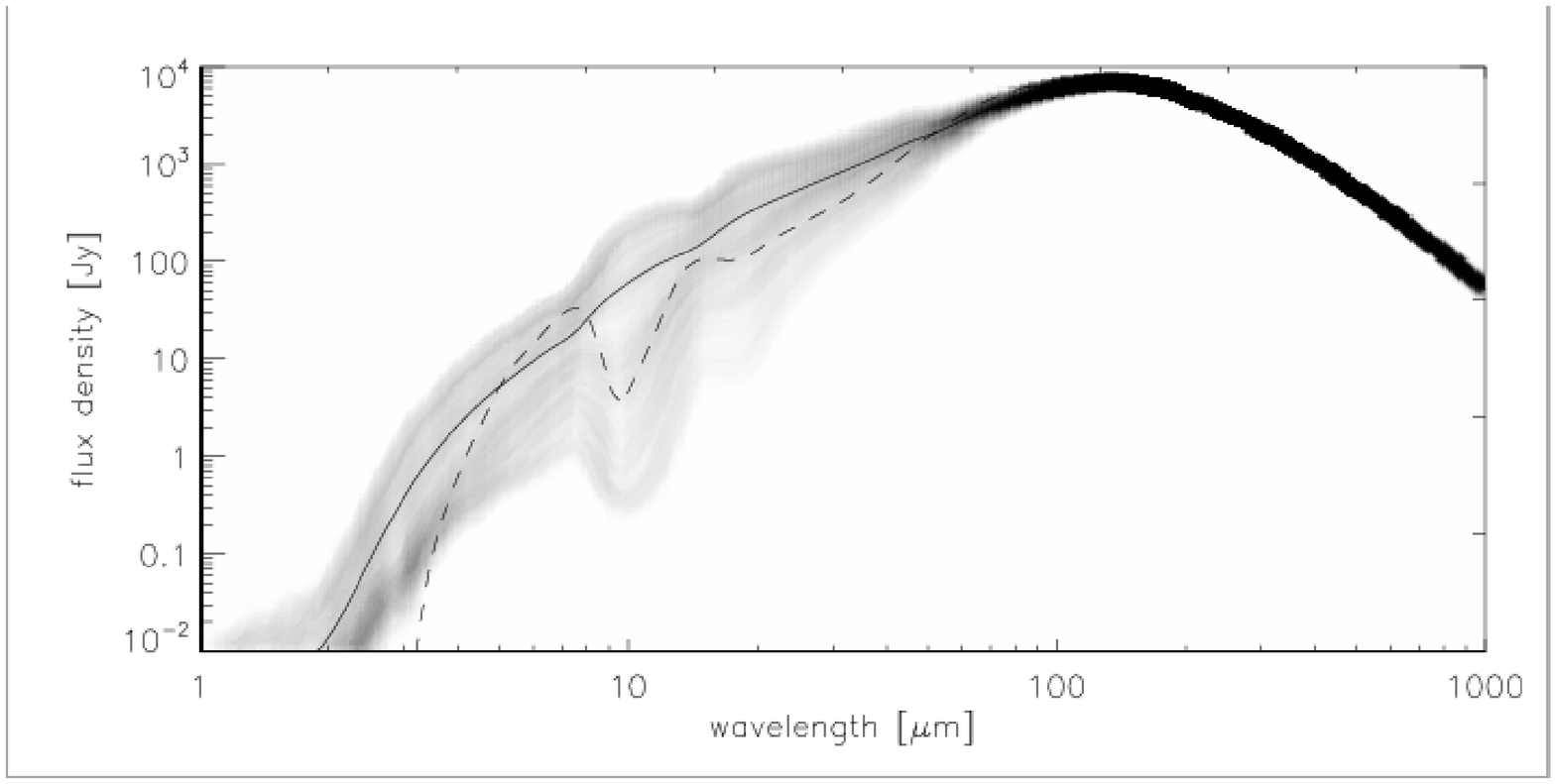}
\rplotone{0.65}{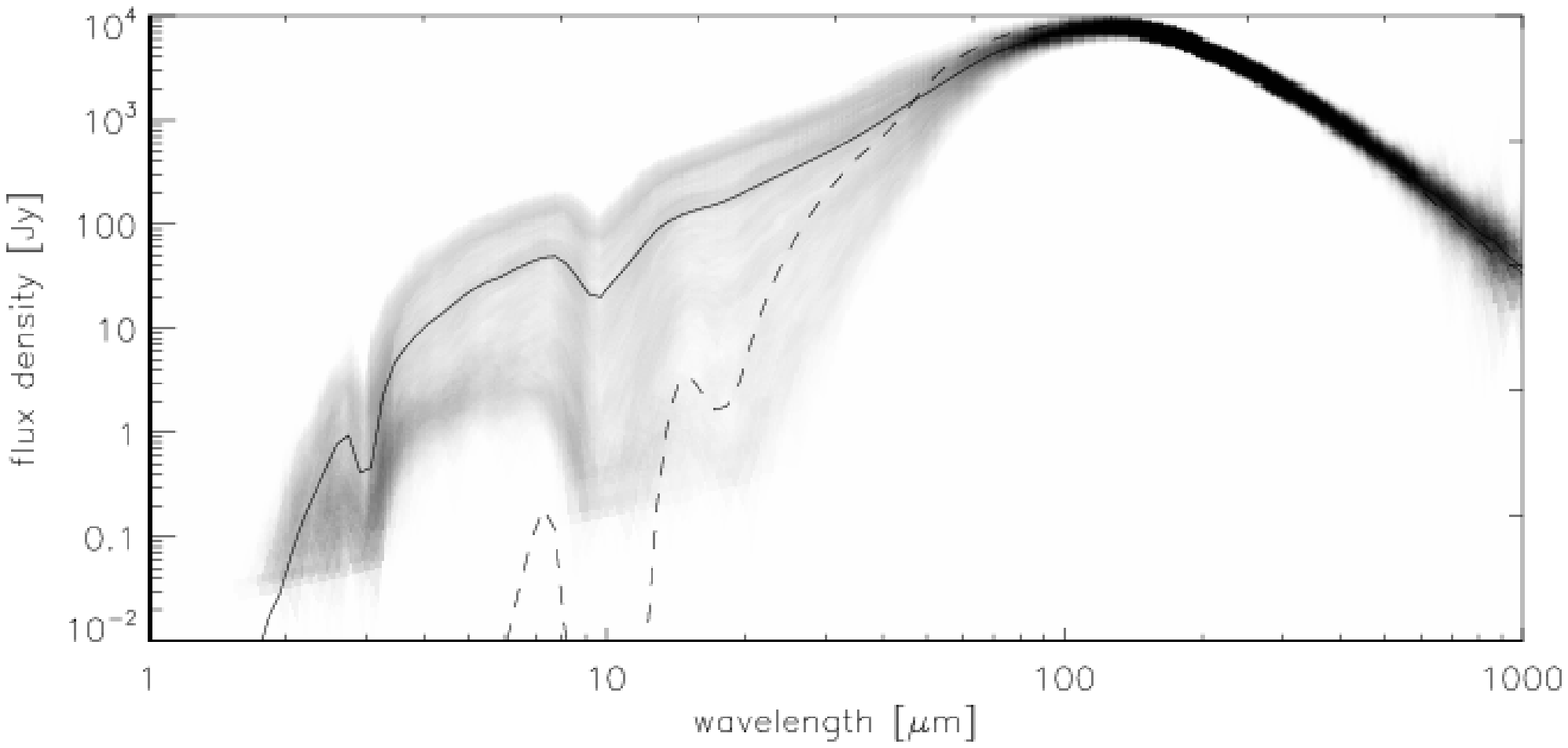}
\caption{\small\small\label{allsedfig} SEDs of clumpy and smooth models.  The
first panel shows the range of SEDs for different sightlines of the
canonical clumpy model.  The grey scale is darker where more similar
SEDs coincide.  SEDs of two particular sightlines (``A'' and ``B'',
see Figs.~\ref{tempdens} and \ref{3col}) are highlighted.  The second
panel compares the clumpy model to a smooth model (dashed line) with
the same dust mass and inner and outer radii.  The average sightline
of the clumpy model is shown as a solid line.
The third panel makes the same comparison between a clumpy and smooth
model, but now with an average radial density power-law gradient
($\rho \propto r^{-1}$).  For simplicity we mostly discuss the
differences between smooth and clumpy models with no radial dependence
of the average density.  In the model with a gradient, the differences
are even larger - both models have higher average optical depth (than
the models without a radial gradient) and more hot dust close to the
star emitting in the NIR, but that emission only escapes and is
observable in the clumpy model, which has low-density sightlines.  In
the smooth model all of that extra emission is preprocessed further
out in the envelope.  }
\end{figure}

We show here a comparison of SEDs of the clumpy and smooth model for
our canonical parameter set discussed in \S~\ref{models}.  As
Figure~\ref{allsedfig} shows, our clumpy model differs from the smooth
model most significantly in the NIR and MIR.  The voids between the
clumps (Figure~\ref{tempdens}) allow NIR flux to scatter through and
escape the envelope.  MIR flux also can escape along certain
sightlines.  The 10 $\mu$m silicate feature varies greatly between the
different sightlines of the clumpy model, but the average depth is
lower than that of the smooth model.  The slope of the SED in the NIR
and MIR is flatter on average in the clumpy model than the smooth.  As
we will show in later sections, these two features of the clumpy
models agree better with observational data of several \uchii\
sources.
\label{gradient}

Figure~\ref{tempdens} showed a slice of the density and temperature
structure for a typical clumpy model, with two particular sightlines
marked.  The SEDs for all 200 sightlines in this model are shown in
Figure~\ref{allsedfig} (in gray), with the two sightlines highlighted.
The output spectra are binned over a finite range in angle, indicated
in Fig.~\ref{tempdens} for each sightline by two solid lines
bracketing the dotted line.  It is important to remember that although
we have only shown the sightline intersecting the central source, the
SED at each viewing angle is due to emission in that direction from
the entire object, i.e. all rays parallel to the range of angles
indicated.
The SED with lesser near-infrared flux corresponds to the lower
(``A'') of the two sightlines in Fig.~\ref{tempdens} -- there are
clumps in the outer envelope in this direction that obscure the inner
regions (this cooler dust also causes the 10$\mu$m silicate absorption
present in the SED, Fig.~\ref{allsedfig}).  Along the upper sightline
``B'', there are few clouds at a radius greater than 0.4~pc, and much
more NIR radiation can scatter out from the inner regions (dashed SED,
Fig.~\ref{allsedfig}).  The orientation of the two sightlines is
particular to this model -- there is no preferred direction to the
clump distribution in general.

\subsection{Synthetic Images}
\label{results}

\begin{figure}
\rplotone{0.8}{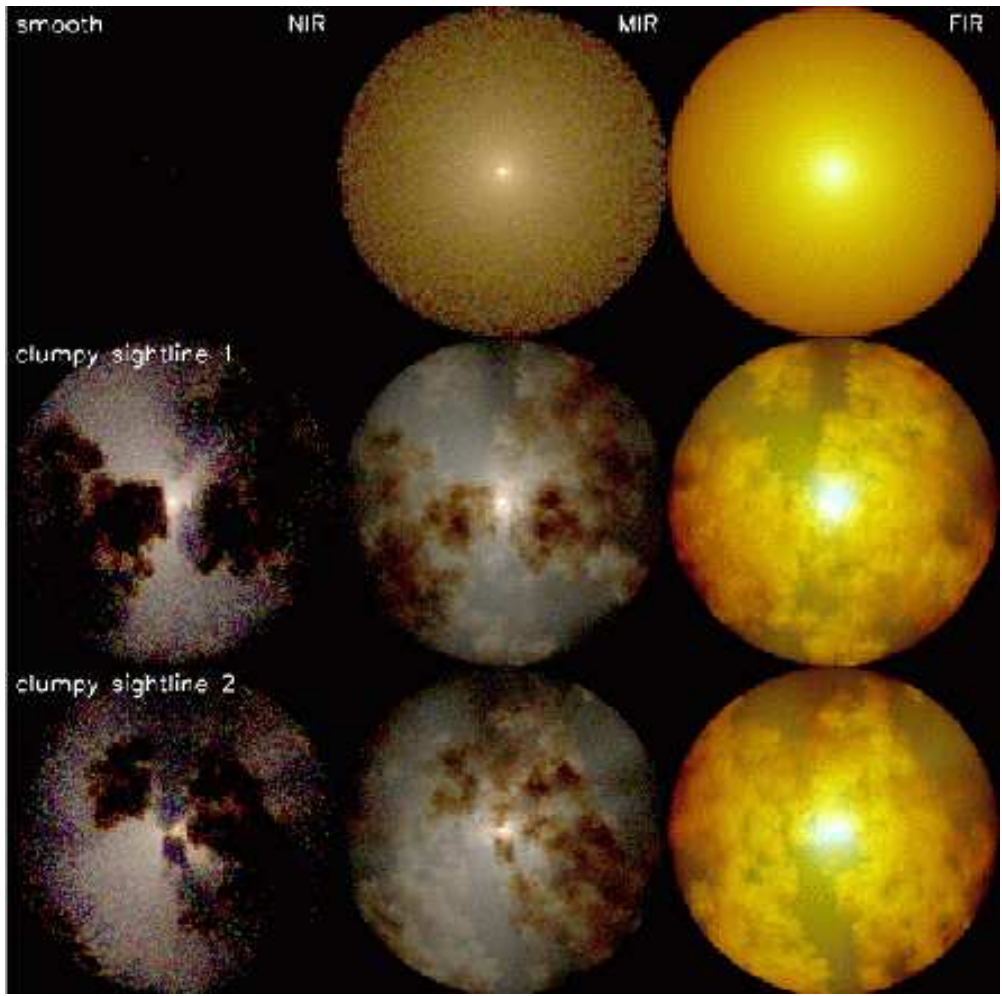}
\caption{\small\label{3col}Synthetic 3-color images of a massive star
embedded in clumpy cloud.  The top row is a spherically symmetric
model, and the second two rows are two views of a clumpy model with
the same mass of circumstellar dust.  (The two sightlines are those
marked in Figure~\ref{tempdens}.)  The NIR (left panels) images are
composed of the emergent spectrum convolved with the 2MASS J,H, and
K$_{\rm s}$ filters as blue, red, and green respectively and a
logarithmic stretch.  The MIR (center panels) images are similarly
composed from synthetic IRAC [3.6], [4.5], and [8.0] images, and the
FIR (right panels) images from synthetic MIPS [24], [70], and [160]
images.  We have not convolved these images with {\it Spitzer}
resolution (1.5\arcsec at 6$\mu$m and diffraction limited at higher
wavelengths); the effect would be negligible for the NIR and MIR
image, but would begin to strongly affect the FIR image at the assumed
distance of 5~kpc.}
\end{figure}

Figure~\ref{3col} shows 3-color images of our canonical clumpy model
as it would be imaged with a standard ground-based NIR camera, and
with the IRAC and MIPS on {\it Spitzer}.  Specifically, the images are
constructed by convolving the simulated output radiation field with
the (J,H,K$_{\rm s}$), ([3.6],[4.5],[8.0]), and ([24],[70],[160])
filters.  The top row shows a spherically symmetric model, and the
second two rows are a clumpy model viewed from different angles.
Clearly, the greatest difference is in the near-IR -- the clumpy model
is $>$6 orders of magnitude brighter at these wavelengths.  The clumpy
model is also brighter in the mid-infrared (IRAC wavelengths), but
more striking is the very different color -- the smooth model is much
redder, with higher average extinction than the clumpy model, and
cooler dust farther out in the cocoon on average.  Differences are
less significant at longer wavelengths, which mostly trace the total
dust mass, and are less sensitive to its distribution.  This agrees
with the SEDs shown in Figure \ref{allsedfig}.

\subsection{Integrated SED}
\label{viewing}

\begin{figure}
\rplotone{0.7}{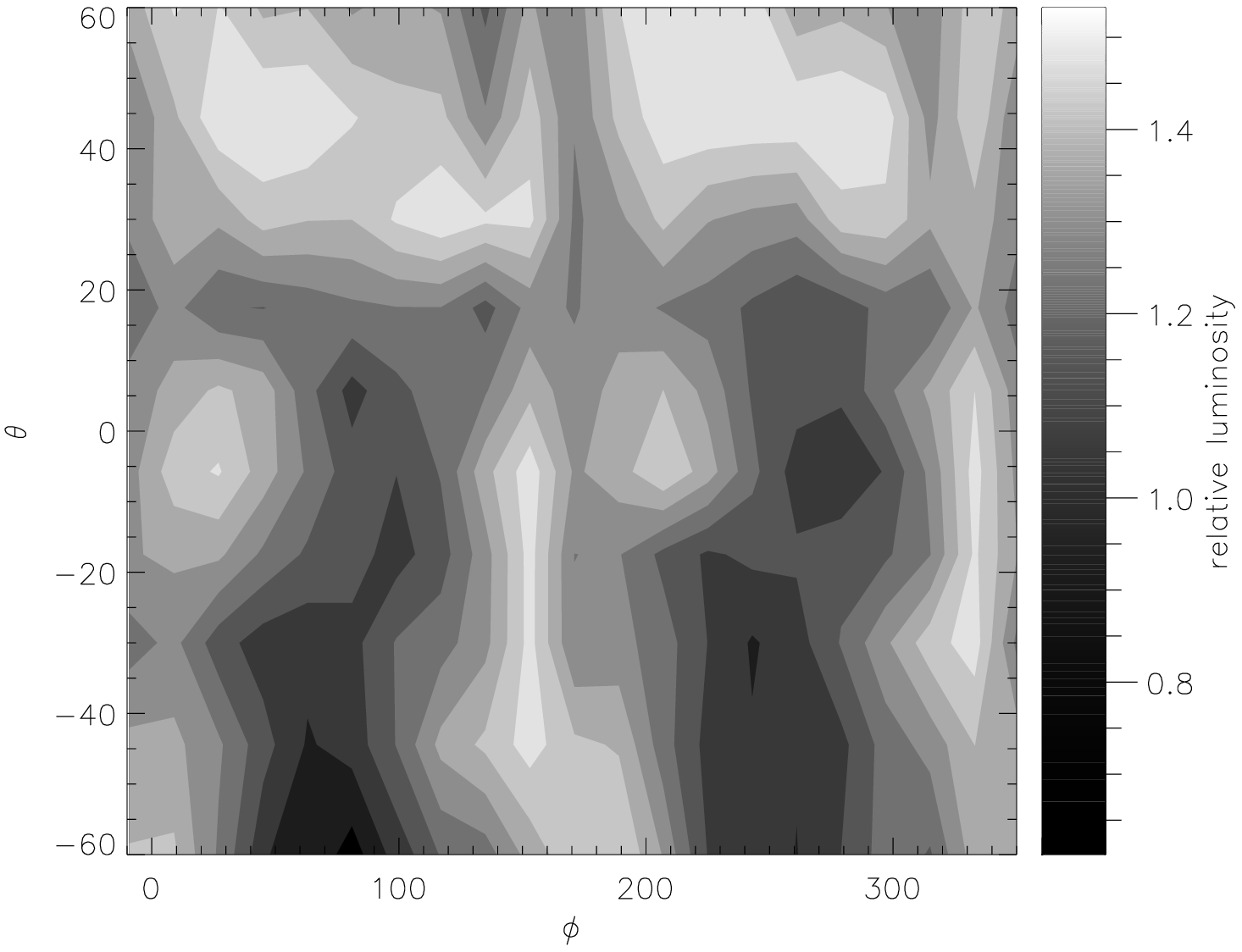}
\caption{\small\label{bolomap} Observed bolometric flux as a function of
polar/altitude ($\theta$) and azimuthal ($\phi$) viewing angle.  The light
spots are angles along which the observed luminosity (if the observer
assumes isotropic emission) is twice the true value, and in the dark
regions the observed value is 60\% of the true value.}
\end{figure}

Not only does the shape of the spectral energy distribution change
with viewing angle, but the total wavelength-integrated flux can
change significantly as well.  Figure~\ref{bolomap} shows contours of
the bolometric flux as a function of $\phi$ and $\theta$, the
azimuthal and polar/altitude viewing angles, for a typical clumpy
model.  The luminosity that an observer would calculate by
(incorrectly) assuming isotropic emission varies from 60\% of the true
luminosity along some sightlines to nearly twice the true value along
other sightlines.  The ``dim'' sightlines are those along which a
dense clump is located at a moderate to large distance from the
central source.  Dense clumps along one's sightline, but very close to
the central source, do not necessarily result in an overall dimming,
because short-wavelength light can scatter around the clump, the
viewer may still be able to see uneclipsed hot dust in the cocoon.  We
note that this variation of the bolometric flux was also found in the
low-mass protostar model of \citet{pI}.
\label{lbol}

\subsection{Colors}

Observationally, the SED of young stellar objects is usually only
sampled with a few broad-band filters, so it is important to show how
the variation apparent in Figure~\ref{allsedfig} translates into a
range of photometric colors.  With the successful launch of {\it
Spitzer}, the availability of 2MASS, and widespread access to deep
near-infrared imaging, photometry of a great number of young stellar
objects is available in the 10 filters $J$, $H$, $K$, IRAC [3.6],
[4.5], [5.8], and MIPS [24], [70], [160], so we concentrate on these
(though we note that in many cases the 160~$\mu$m band is not very
observationally useful because of its low resolution and saturation
limit).  For conversion of fluxes to magnitudes we use 1594., 1083.,
and 666.7~Jy for 2MASS $J$, $H$, and $K_s$, 277.5, 179.5, 116.6, 63.1,
and 7.16~Jy for the four IRAC bands and MIPS [24] (M.~Cohen 2004,
private communication), and an estimated 0.75 and 0.2~Jy for MIPS [70]
and [160].  We also calculate the IRAS fluxes of our models and
discuss these briefly in \S\ref{wccolors}.

\begin{figure}
\plottwo{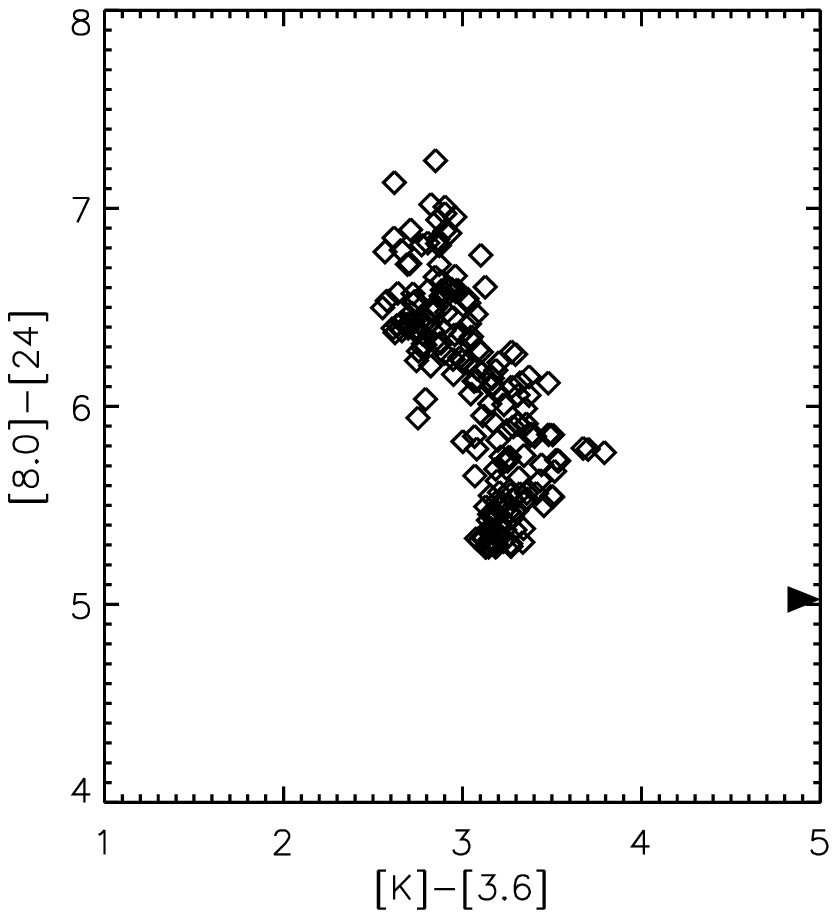}{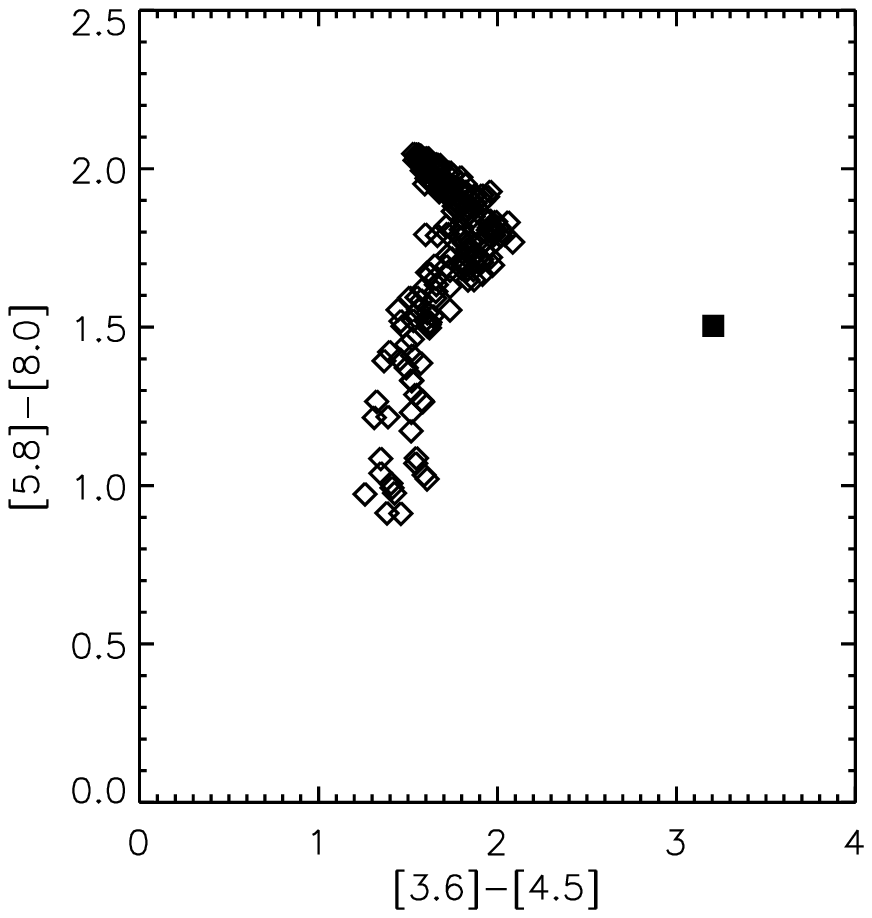}
\caption{\small\label{ccdangle} Color-color diagrams (CCDs) for different
observations of a {\it single} modeled object, an 0-type star with
50000~M$_\sun$ of circumstellar material distributed in clumps between
0.0001 and 2.5~pc radius (the same model as
Figs.~\ref{tempdens}-\ref{bolomap}).  The different points are for the
200 different viewing angles calculated for each model.  In the first
panel, colors across the broad wavelength range 2--25$\mu$m are shown,
and in the second panel colors derivable exclusively from IRAC
photometry.  The corresponding smooth model is shown as a filled
triangle in the first panel, indicating that the actual point is far
off the plot to the right, and as a filled square in the second
panel.}
\end{figure}

Figure~\ref{ccdangle} shows two color-color diagrams for our example
typical clumpy model, viewed from 200 different angles.  The
near-infrared bands vary the most with viewing angle, but colors
formed with an IRAC band and MIPS [24] are also quite sensitive to the
circumstellar dust distribution.  This is because the MIPS bands are
at long enough wavelengths that they sample predominantly thermal
emission of fairly cool dust; the flux density mostly depends on the
total dust mass and is less affected by the particular distribution.
We also show the color-color diagram formed exclusively from IRAC
bands, since many young stellar objects in the inner Galaxy are being
imaged with IRAC by the GLIMPSE survey \citep{pasp,edsup,barbsup}.
Colors can vary with viewing angle by more than a magnitude for the
same object.  The corresponding smooth model (same dust mass and inner
and outer radii) is shown on the two CCDs, indicating that even though
the variation between sightlines in the clumpy model is large, the
difference between any sightline and a smooth model is even larger.
The smooth model is extremely embedded and nearly invisible in the
NIR, so the [K]-[3.6] color is far off the right side of the first
panel. Even the IRAC-only colors of the smooth model are different
from the clumpy model by more than a magnitude.

\subsection{Fitting 3-D data with 1-D Models}
\label{ave}

\begin{figure}
\rplotone{1.}{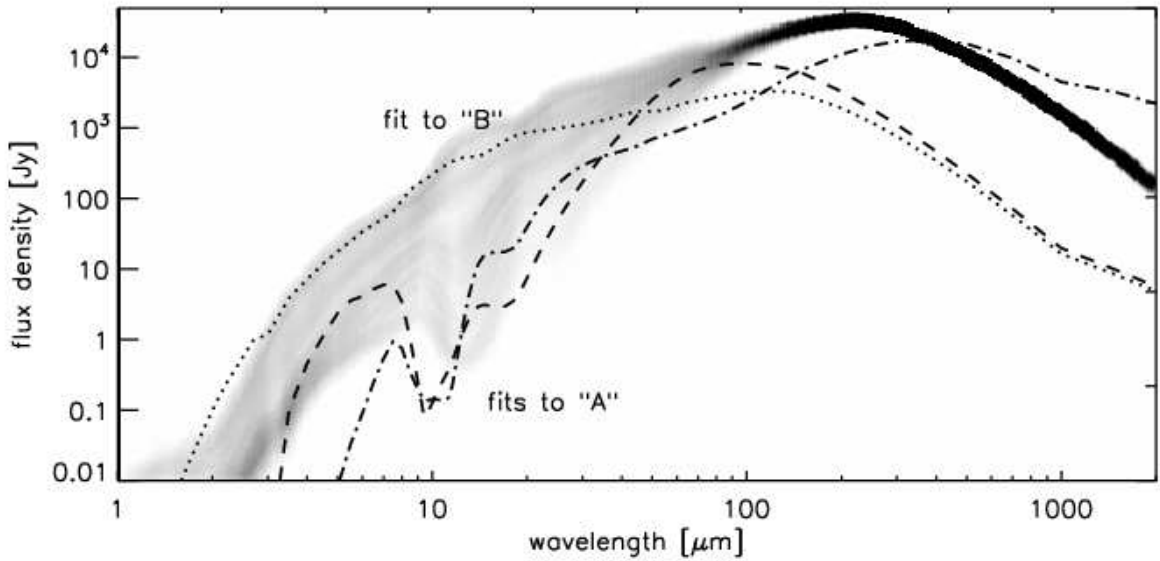}
\caption{\small\label{fitrange}Attempted 1-D fits to the SEDs of a 3-D
clumpy model.  The variation of SEDs with viewing angle is shown in
grayscale as in Figure~\ref{allsedfig}.  The high and low extremes of
the greyscale correspond to the two sightlines shown in
Fig.~\ref{tempdens} that have particularly high and low NIR flux
densities (``B'' and ``A'', respectively, see first panel of
Fig~\ref{allsedfig}).  The dotted line is the best 1-D fit to SED
``B''.  The dashed and dot-dashed are 1-D fits to SED ``A''. The 1-D
model parameters do not reflect the true dust distribution very
closely -- see text for more discussion.}
\end{figure}

The spectral energy distributions of 3-D clumpy models are
significantly different from those of 1-D smooth models.  We attempt
to fit the output of our 3-D models with 1-D models, revealing 
several specific differences: 1) The derived physical
parameters (dust mass, etc) are often very different, 2) the derived
physical parameters for the same object depend on viewing angle, and
3) sometimes it is impossible to fit the (synthetic) data with a 1-D
model at all.  Figure~\ref{fitrange} illustrates these points.  The
range of SEDs for a single clumpy model seen from different angles is
shown as a gray scale, along with three 1-D models.  The ``true''
physical parameters are 50000~$M_\sun$ of circumstellar material
around an O-type star (T$_\star=$41040~K,
L$_\star=$2.54$\times$10$^5$~L$_\sun$).  Some very thin dust extends
all the way in to 0.0001~pc (the density is never truly zero), and the
closest clump (typical density of clumps is 10$^5$~cm$^{-3}$) is at
radius 0.035~pc, and temperature 250~K. The average optical depth to
the central star is $A_V=$131, but that quantity ranges from 13
to 401 magnitudes depending on viewing angle.

We have concentrated here on fitting the NIR and MIR parts of the SED,
without heavily constraining the FIR.  This is motivated by current
observational capability, which includes the widespread availability
of NIR spectrographs and cameras and {\it Spitzer}'s relatively
high-resolution coverage of 3-30$\mu$m.  IRAS and even the 70 and
160$\mu$m channels of MIPS on {\it Spitzer} are resolution-challenged,
and it is more difficult to draw conclusions about massive protostars
(generally at distances $>$1~kpc and in crowded environments) from
these instruments.  One can instead try to crudely fit the shape of
the entire SED using a 1-D model; a reasonable match to the SED shape
can be made for the least embedded sightlines as shown below. However,
in general these 1-D fits either fail to reproduce the 10$\mu$m
silicate feature in absorption or have MIR ($\sim$5$\mu$m) fluxes that
are 1-3 orders of magnitude too low, and more seriously, the physical
parameters derived are not the same as the 3-D model (size and mass of
the cocoon). 

\begin{figure}
\plottwo{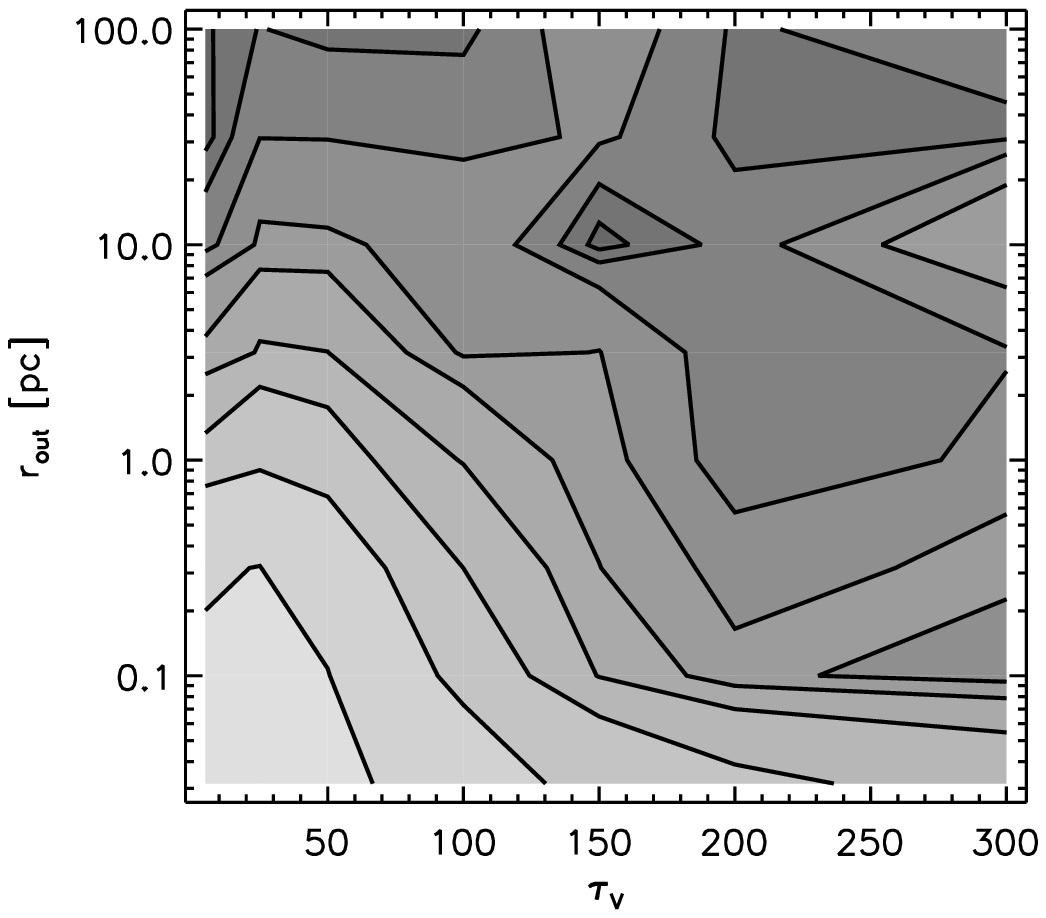}{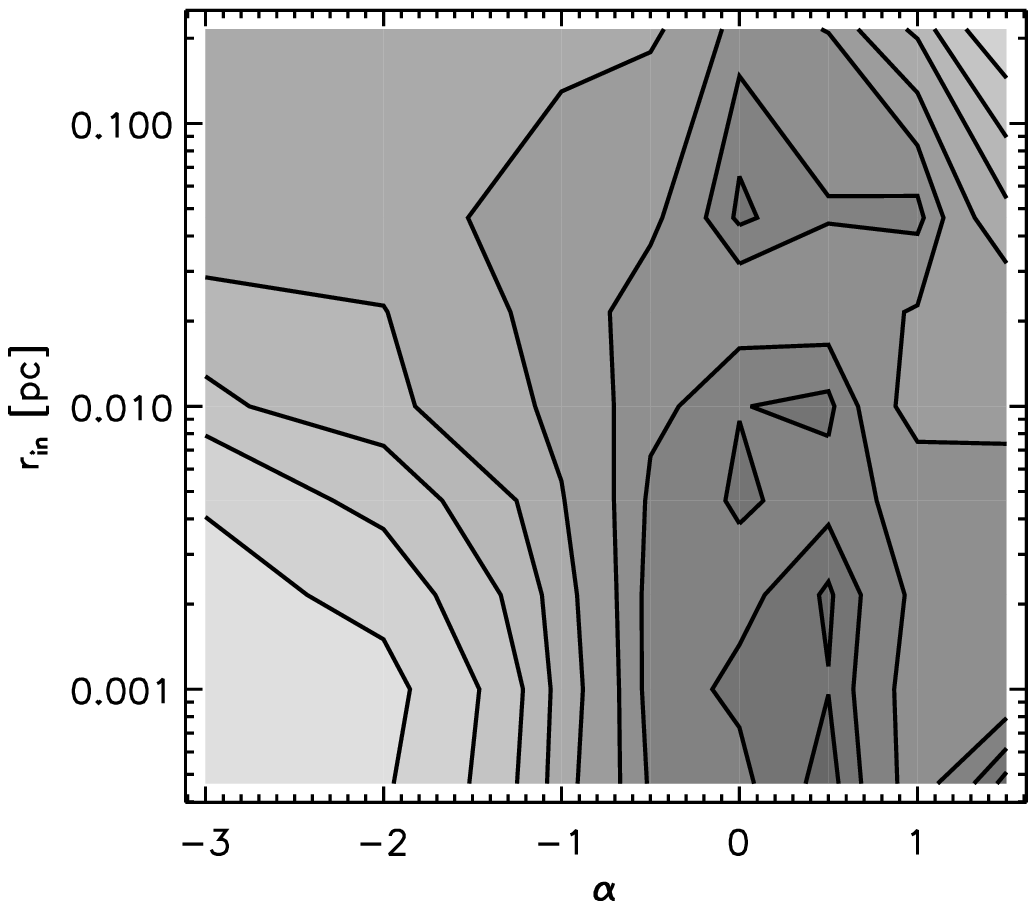}\\
\plottwo{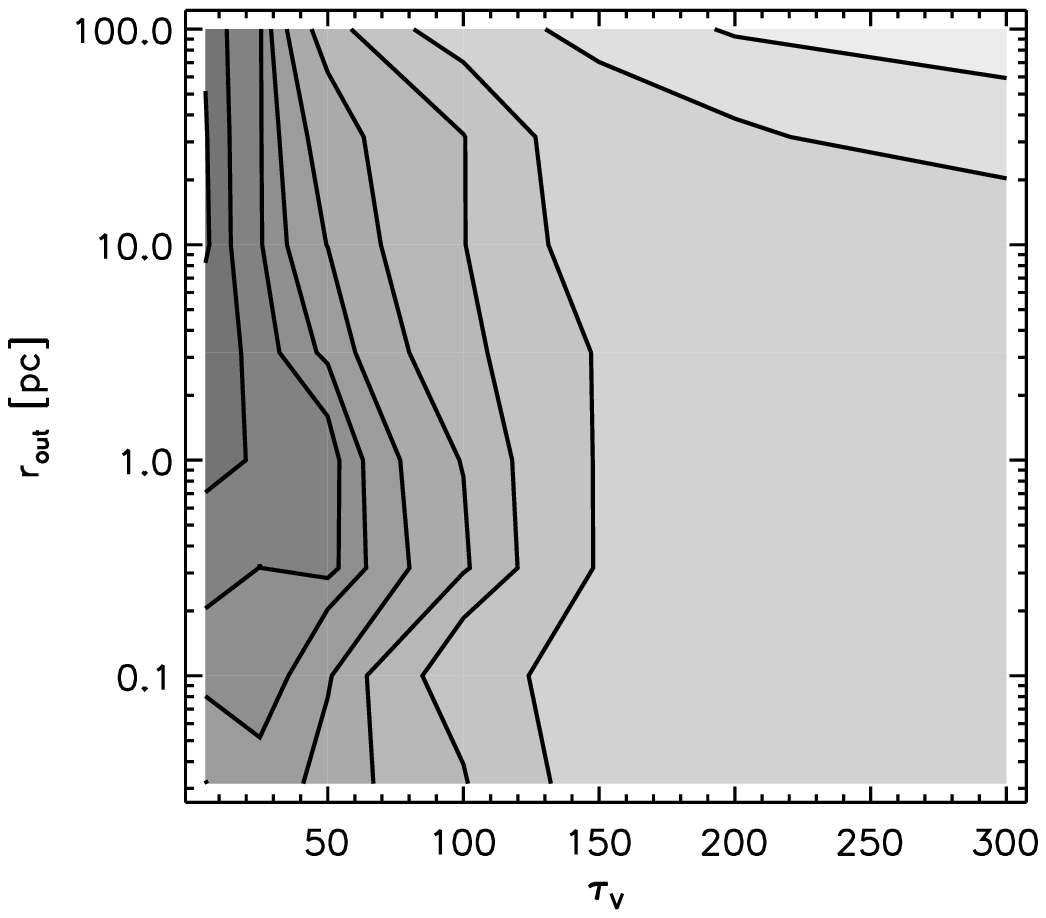}{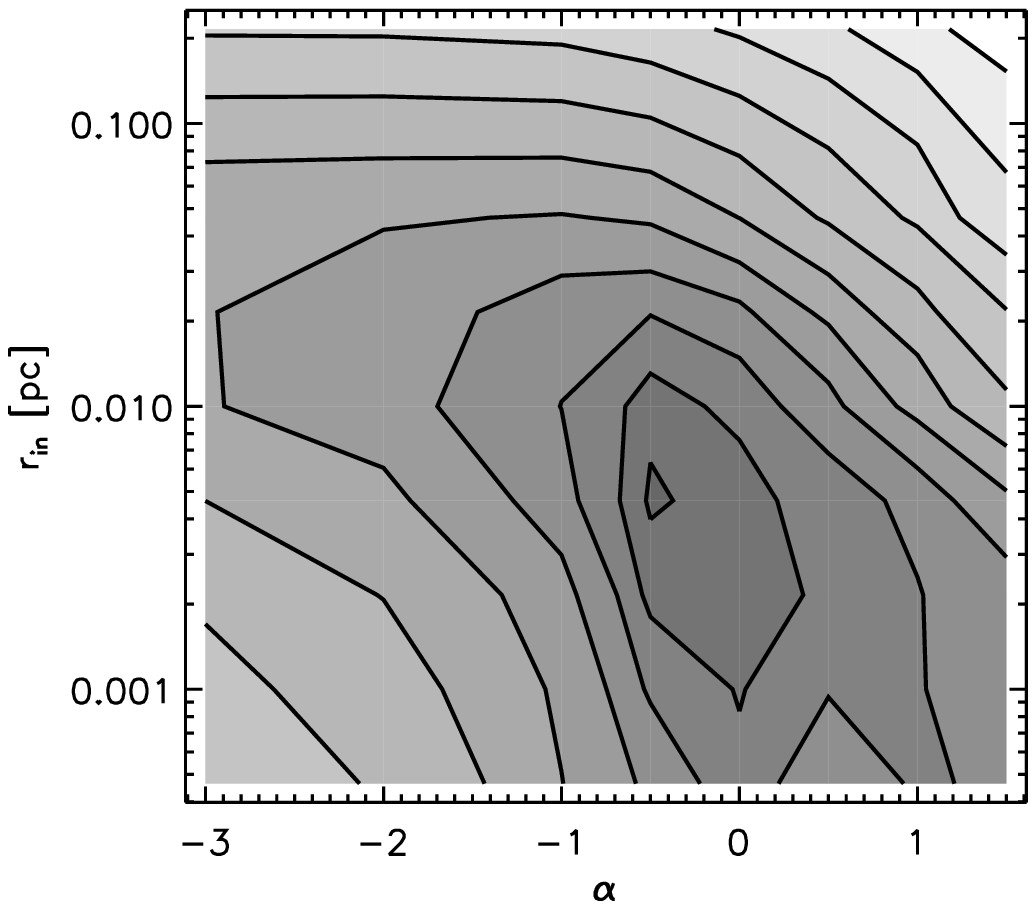}\\
\caption{\small\label{modelcontour}
Goodness of fit of 1-D smooth models to two selected 3-D
clumpy SEDs.  The first two panels represent attempts to fit sightline
``A'' with a strong silicate absoprtion feature.  The second two
panels are for sightline ``B'' which shows silicate emission.  In both
cases, sightlines with fairly flat radial density profiles and larger
inner and outer radii than the true model are preferred. The best fits 
have fairly poor ``$\chi^2$'' values of $5-10$.}
\end{figure}

In order to quantify the differences between the SEDs produced by
smooth 1-D and clumpy 3-D models, we constructed a grid of 5000 smooth
models, varying the inner and outer radius, the optical depth or dust
mass, and the radial density profile (power-law profiles
$\rho(r)\propto r^\alpha$).  We then calculated the ``$\chi^2$''
goodness-of-fit between each 1-D model and a selected 3-D sightline
(and below, observation of a real YSO).  We used an arbitrary
``uncertainty'' of 0.1 dex to calculate $\chi^2$ for the model SEDs --
the preferred parts of parameter space are not affected by this
choice.  Figure~\ref{modelcontour} shows this goodness of fit as a
function of the optical depth $\tau_V$, the inner and outer radii
$r_{in}$ and $r_{out}$, and the radial density power-law index
$\alpha$.  

The dotted line shows the best 1-D fit to the 3-D sightline with
particularly high NIR/MIR flux density (``B'' in Fig.~\ref{tempdens}).
The best fitting models have fairly flat radial density profiles
($\alpha=0.0$; fit parameters are listed in Table~\ref{fittable}).
The preferred optical depths are much lower than the average optical
depth of the 3-D model (21\% for this best-fit model), so the dust
mass would be severely underestimated.  If one decreases the optical
depth in the smooth model in order to increase the flux density at
shorter wavelengths, one is also forced to increase the outer radius
and total dust mass to maintain the observed FIR flux density.  It is
difficult to produce sensible 1-D models that have the
10$\mu$m silicate feature in emission as well as the shape of the SED
at other wavelengths (in the 3-D models this is possible for viewing
angles at which very hot dust near the star is not severely obscured.)

\begin{deluxetable}{lccccccc}
\tablecolumns{8}
\tablewidth{0cm}
\tablecaption{\small\label{fittable} Parameters of 1-D models fitting 3-D models}
\tablehead{\colhead{model} & \colhead{density} & 
\colhead{r$_i$}  & \colhead{r$_o$} & 
\colhead{mass} & \colhead{luminosity} & \colhead{Average ${\rm A_V}$} & \colhead{``$\chi^2$''} \\
& \colhead{index $\alpha$\tablenotemark{a}} & \colhead{[pc]} 
& \colhead{[pc]} & \colhead{M$_\sun$} & 
\colhead{10$^5$ L$_\sun$} & \colhead{to star} & }
\startdata
clumpy                         & 0   & 0.0001\tablenotemark{b} & 2.5 & 50000 & 2.54 & 131 & -- \\
\hline
\multicolumn{7}{l}{1-D fits to lower sightline ``A''}\\ 
dashed line Fig~\ref{fitrange} & 0   & 0.047            & 47. &  2.9$\times$10$^7$ & 2.54 & 217 & 8.7 \\
dot-dash line Fig~\ref{fitrange}&+1.5& 0.00039          & 12. &  2.0$\times$10$^6$ & 2.54 & 163 & 4.8 \\

luminosity unconstrained       & 0   & 0.0005           & 91   & 5.6$\times$10$^7$ & 0.6  & 109 & -- \\
                               & -1  & 0.0070           & 7.0  & 1.4$\times$10$^5$ & 0.5  & 200 & -- \\
\hline
\multicolumn{7}{l}{1-D fits to upper sightline ``B''}\\ 
dotted line Fig~\ref{fitrange} & 0   & 0.0027           & 2.7 &  2.3$\times$10$^5$ & 2.54 & 27  & 8.2 \\
luminosity unconstrained       & 0   & 0.0023           & 3.5 &  31300             & 3.5  & 43  & -- \\
\enddata
\tablenotetext{a}{Smooth models were run with a power-law density 
gradient $\rho\propto r^{\alpha}$.}
\tablenotetext{b}{Clumpy models have diffuse material extending in to
the dust destruction radius, but the distance to the nearest dense
($\gtrsim$10$^5$cm$^{-3}$) clump is farther out (0.035pc).}
\end{deluxetable}

It is even more difficult to fit the sightlines like ``A'' that have
low MIR intensity using a 1-D model.  The dot-dashed line shows the
best fit constant-density model, which has a reasonably deep silicate
absorption feature but not enough NIR flux and too much flux at
20$\mu$m.  This is a general feature with 1-D models that try to mimic
typical sightlines of the 3-D model: the optical depth must be kept
moderate (similar or somewhat greater than the average value of the
clumpy models) to produce the silicate absorption feature, and then
there is no way to get emission shortward of 2$\mu$m out of a
reasonably embedded source without an inhomogeneous envelope.  In 1-D
models the depth of the silicate feature is tied to the mass of the
envelope, but three dimensions these are decoupled.

The best-fitting models also have unreasonably large outer radii, and
often, large inner holes or radial density profiles that increase with
radius (a low-density inner region is similar to a hole of zero
density).  It is these last models (positive $\alpha$) which produce
the peak in the goodness-of-fit at very low $\tau$ and very large
$r_{out}$ -- we do not consider these models physically reasonable, but
plot one for completeness (dashed line).  We note that 1-D fits to 
real data have the same characteristics -- flat 
radial profiles and large outer and inner radii. (Our fits
are discussed below, others such as F98 found the same result.)

If one did not know the intrinsic luminosity of the source, that
parameter would also likely be fit incorrectly.  We did not construct
a full grid of models with different luminosities, but explored a
range of parameter space fitting by eye, and found that the fit could
be improved slightly over the SEDs shown in Figure~\ref{fitrange} by
changing the source luminosity (downward for sightline ``A'' and
upward for sightline ``B'' -- those model parameters are also listed
in Table~\ref{fittable}).  We will see in later sections that 1-D
models face similar difficulties in fitting real data that they do in
fitting 3-D SEDs.


\section{More General Clumpy Models -- Observable Trends}

\subsection{Relating the SED and the dust distribution}
\label{trends}

Although the same young stellar object can look different depending on
the viewing angle, trends do exist that are strong enough to overcome
that intrinsic scatter and probe the physical state of the
circumstellar dust.  Spatially averaged information about the density
distribution and size of the inner cavity, the magnitude of density
fluctuations relative to the mean, and the total dust mass are all
observationally accessible.

\subsubsection{Trends with average $A_V$ and inner cavity size}
\label{radtrendsec}

\begin{figure}
\rplotone{0.7}{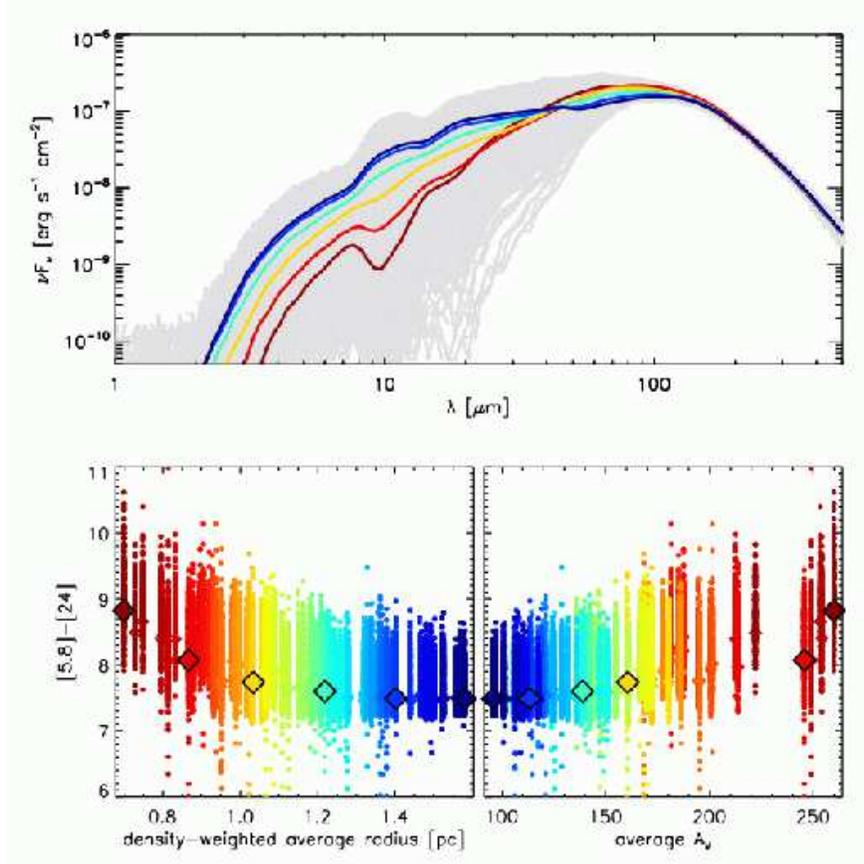}
\caption{\small\label{radtrend} One hundred clumpy models with the same
amount of circumstellar mass in different random clump
distributions.  The upper panel shows the range of the 20000
sightlines (200 per model) as gray lines.  Five particular models have
been highlighted in color, and the {\it angle-averaged} SED plotted
for those five.  The lower panels show how the 5.8$\mu$m-24$\mu$m
color varies as a function of the density weighted average radius (a
measure of the distance from the central source at which most of the
dust is located) and of the angle-averaged optical depth to the
central source. The 200 sightlines for each model are marked as dots
in the lower panels, and the angle-averaged properties of the five
highlighted models are shown as large diamonds. }
\end{figure}

One intuitive fact about circumstellar dust that remains true even in
the presence of complex density distributions is that as the optical
depth to the source of emission increases, the emergent near- and mid-
infrared light decreases and the source appears redder.  If one
considers different sources with the same mass of circumstellar dust,
then the size of an inner cavity or hole should be anticorrelated with
the optical depth to the central source.  In clumpy models, there is
not a spherical inner hole, but the average distance of dust from the
central star varies depending on the specifics of the clump
distribution.  We can explore this effect in a natural way by running
many models with different random seeds or realizations of the clump
distribution.  Figure~\ref{radtrend} illustrates the results: We show
100 models with a smooth-to-clumpy ratio of 0.1 (mostly clumpy), and
different random seeds of the clump distribution.  The 20000
SEDs (200 viewing angles per model) vary widely at the shorter
wavelengths, and the [5.8]-[24] color shown in the lower panels can
vary by over two magnitudes for a single model.  The expected trend is
nevertheless visible - models in which the clumps are largely further
from the central source have somewhat lower average optical depth, and
are somewhat bluer.  The trend is quite clear in the angle-averaged
properties of models (see the five highlighted models in particular),
but the angle-averaged properties are not observable from Earth.

\begin{figure}[h]
\rplotone{0.6}{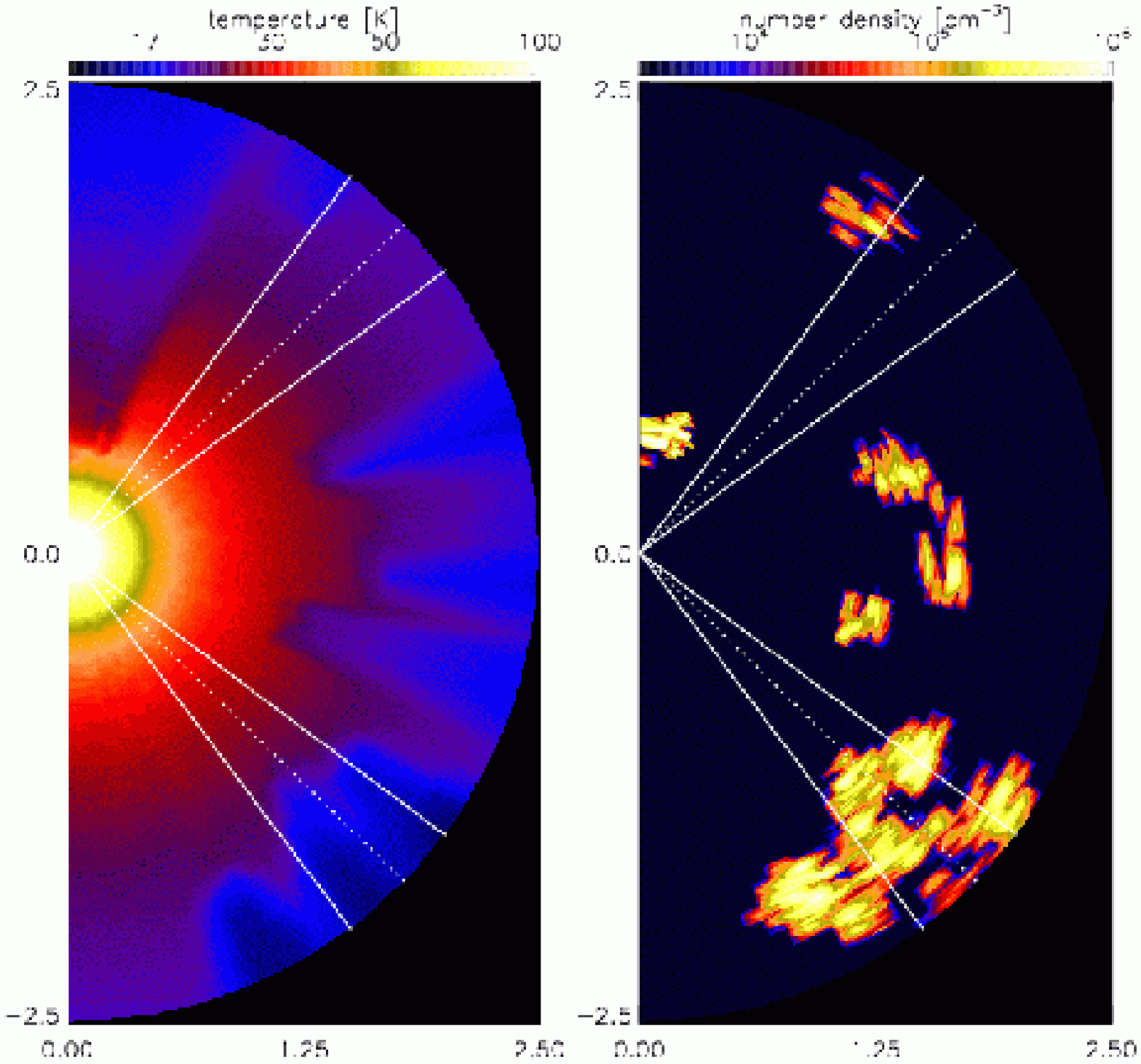}
\rplotone{0.6}{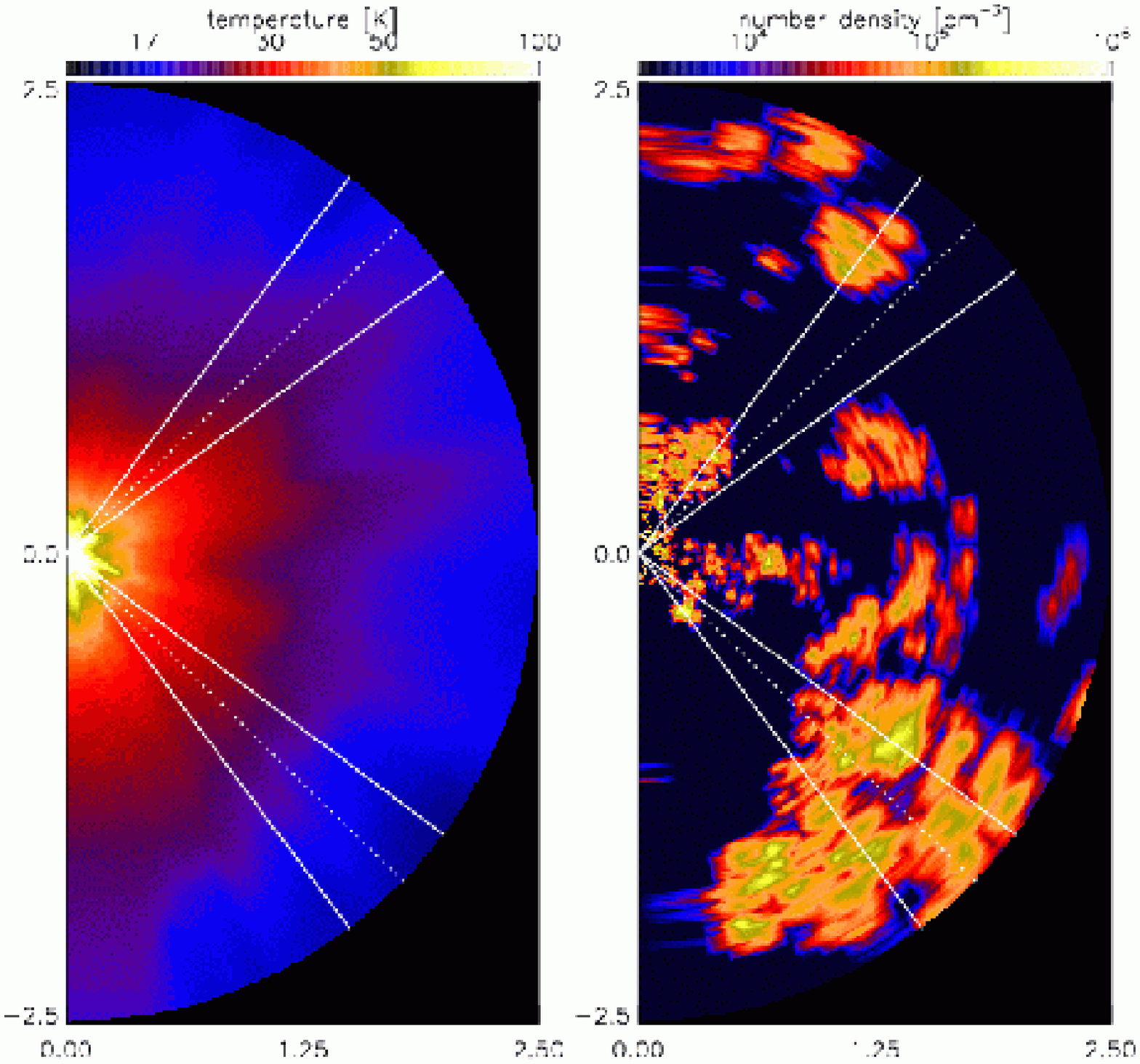}
\caption{\small\label{twofract} Models with effective fractal dimensions
$D$=2.4 and 2.9 - compare to Fig.~\ref{tempdens}, which has $D$=2.6,
the standard used in most of our models.  Higher effective fractal
dimension corresponds to ``fluffier'' clouds, a more filled cloud
distribution, and lower density peaks.}
\end{figure}

\begin{figure}[h]
\rplotone{0.7}{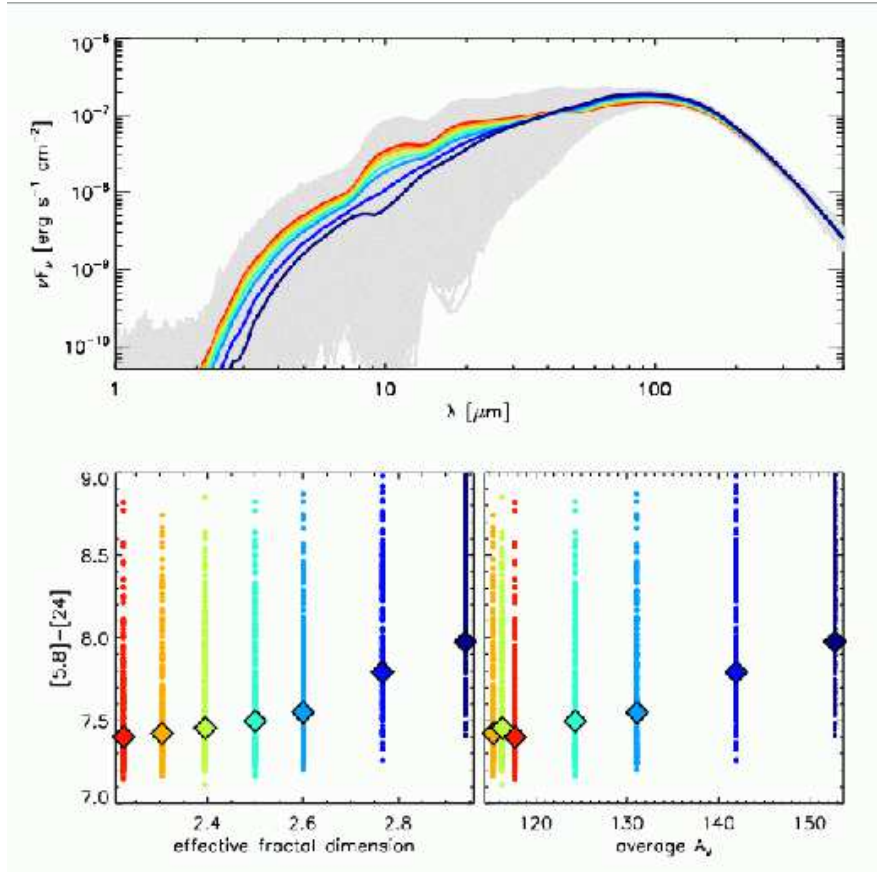}
\caption{\small\label{fracttrend} The effective fractal dimension $D$ does
not have a strong effect on results (within the range shown, which is
believed to be the relevant range for the ISM).  The 200 sightlines
for each of 7 models with different $D$ are shown as gray scale in the
upper panel, and dots in the lower panel.  The angle-averaged
properties are shown as thick SEDs and large diamonds.  The variation
in observable properties with $D$ is less than the variation of a
single model with viewing angle. }
\end{figure}

\label{fracdimsec}
The effective fractal dimension $D$ offers another way to quantify the
dust distribution -- a larger effective dimension corresponds to more
``fluffy'' clouds with a larger filling factor.  Most of our models
use $D$=2.6, close to the effective dimension suggested by
\cite{clumpyISM} for the general interstellar medium, but we verify
here that within a reasonable range, this choice does not have a
strong effect on our results.  Figure~\ref{twofract} shows slices of
models with higher and lower $D$, as compared to the standard shown in
Fig.~\ref{tempdens}.  The SEDs and properties of models with a range
of $D$ are shown in Figure~\ref{fracttrend}, where as before the full
range of sightlines is shown and selected angle-averaged properties
are highlighted. The variation of observable properties (SED, colors)
is less than the variation as one changes the viewing angle for a
single model.  A fluffier or more filled cloud distribution results in
somewhat redder colors and larger average optical depth, which is not
surprising, but the effect is smaller than others examined in this
paper, such as the cloud/intercloud density contrast (smooth-to-clumpy
ratio).


\clearpage

\subsection{Trends with smooth-to-clumpy ratio}

\begin{figure}
\rplotone{0.65}{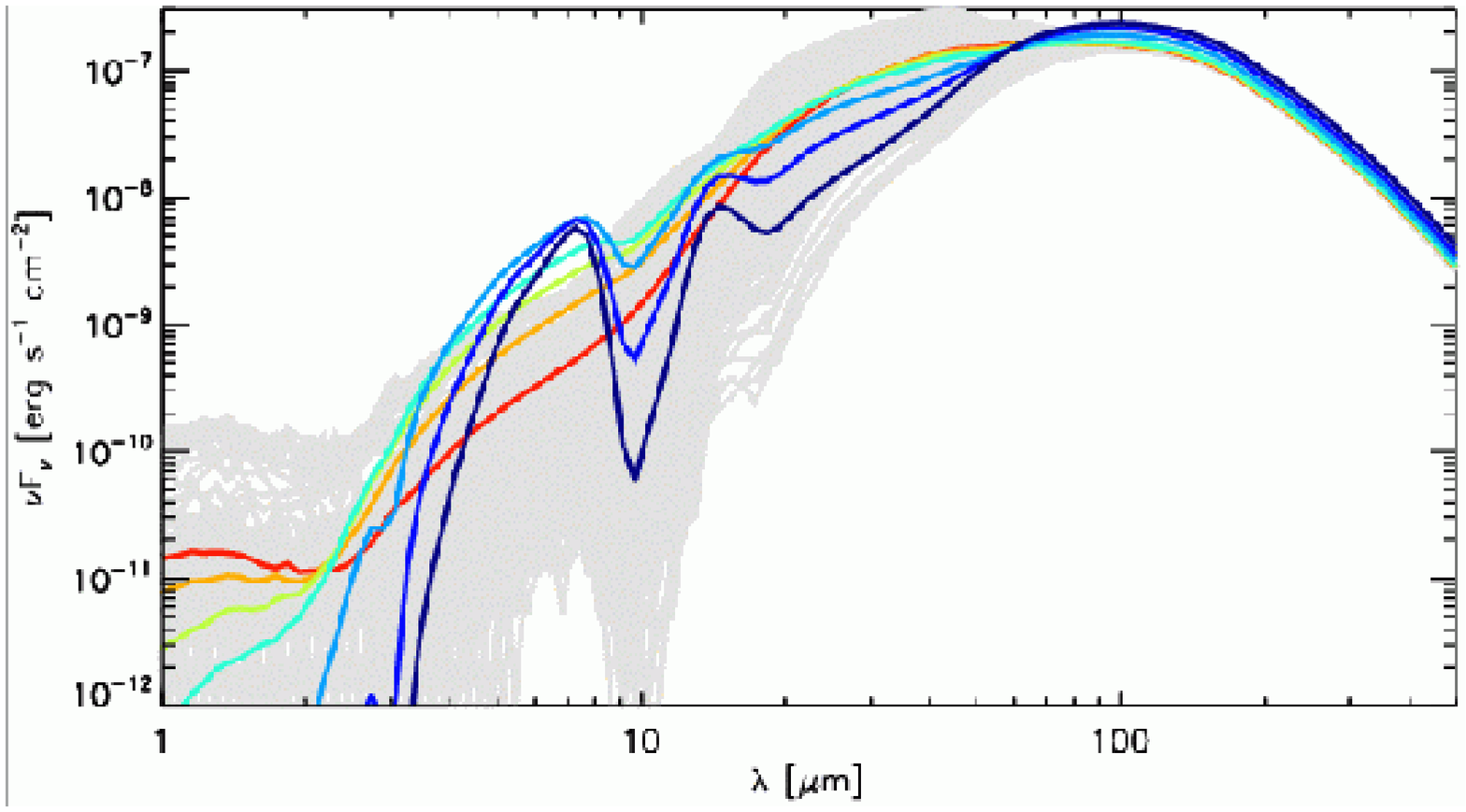}
\rplotone{0.65}{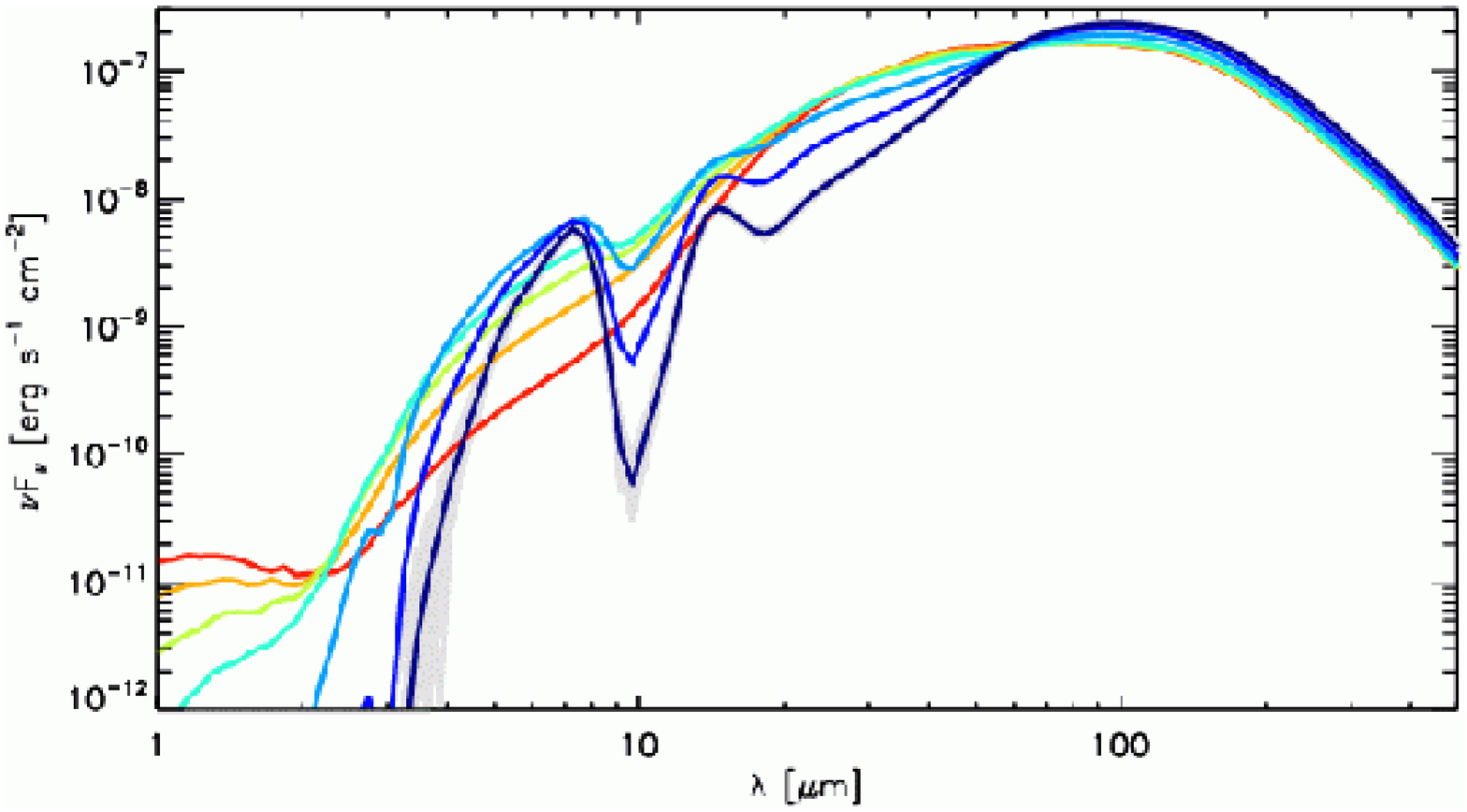}
\caption{\small\label{plotdensrat} SEDs for models with different
ratios of smooth to clumpy density distribution, from completely
clumpy (lightest gray; red in color version) with nearly evacuated
voids to completely smooth (darkest grey; dark blue in color version).
Each colored line in the upper plot represents the average of 200
viewing angles, and the gray scale shows the range spanned by those
individual sightlines (as the average sightline is not observable from
earth).  In the upper plot the gray scale is the range spanned by the
model with 1\% smooth-to-clumpy ratio (very evacuated voids), and in
the lower plot the gray scale is for the 90\% smooth model.  Although
there are still large variations with viewing angle, the smoothness of
the cocoon is a large enough effect to be observationally probed.}
\end{figure}

\begin{figure}
\rplotone{1.}{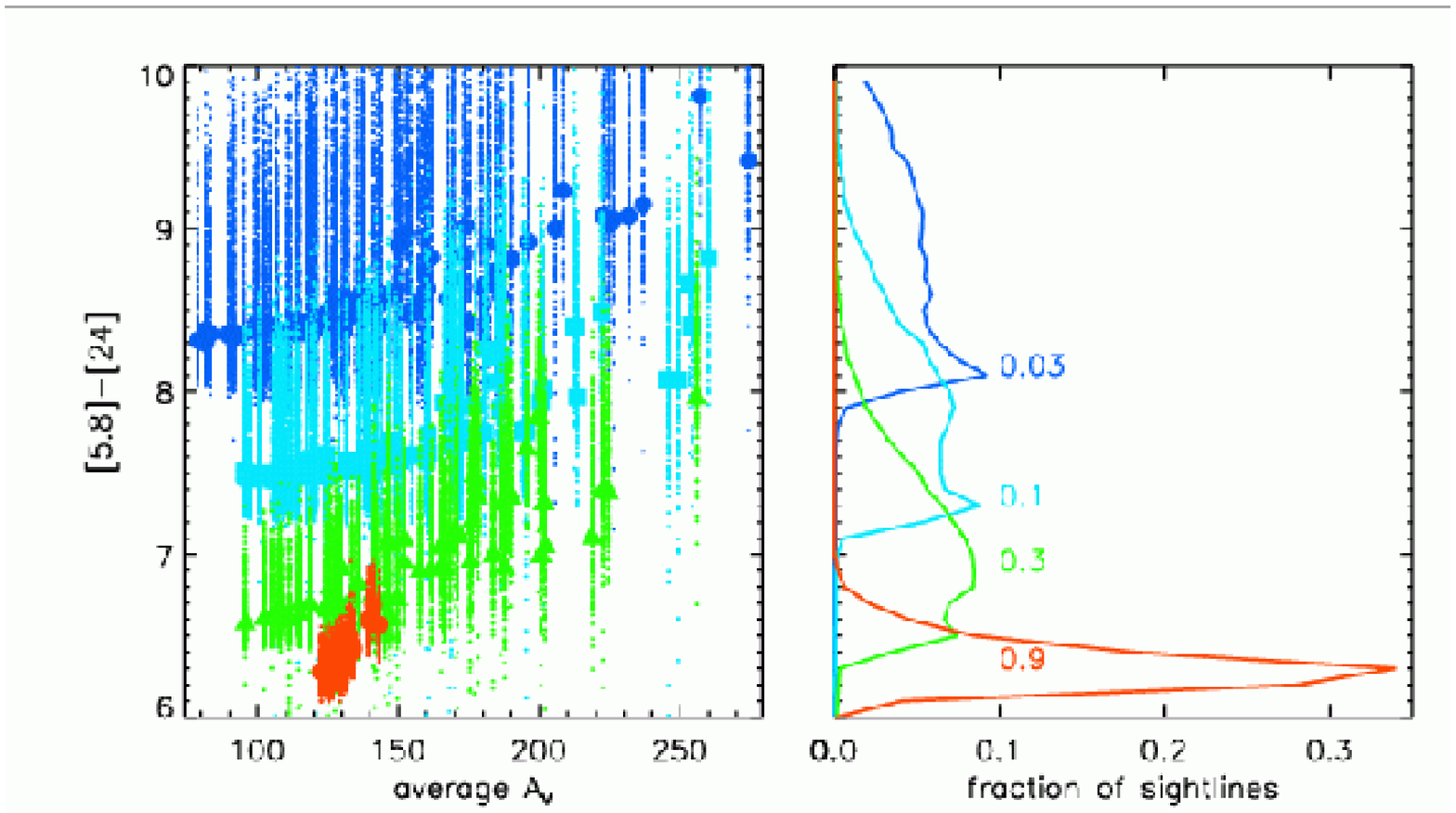}
\caption{\small\label{denshist} The left panel of this figure is similar to
the lower right panel of Fig.~\ref{radtrend}, showing 244 models with
different distributions of the same total mass of clumpy clouds, but
here different ratios of smooth to clumpy material are shown, 0.03
(blue circles), 0.1 (squares), 0.3 (triangles), and 0.9 (red circles).
The right panel collapses the horizontal axis, showing histograms of
the observable quantity, [5.8]-[24] color, labeled according to
smooth-to-clumpy ratio.  Clearly, the color can diagnose the degree of
clumpiness of the medium (if the luminosity of the central source is
known and the total mass can be estimated e.g. from FIR
observations).}
\end{figure}

To quantify the effects of a clumpy circumstellar medium in another
way, we varied the ratio of the typical clump density to that of the
smooth interclump medium.  Figure \ref{plotdensrat} shows the effect
of varying the smooth-to-clumpy ratio for models the same particular
random distribution of clouds.  As in similar figures above,
we show the range spanned by different viewing angles as well as the
average properties to highlight the trend.  As the smooth-to-clumpy
ratio increases, the depth of the 10$\mu$m silicate feature increases,
and the models become redder shortward of that feature.  The MIR flux
actually increases slightly for some parts of parameter space: the
clumpy models can have a central hole with fairly thin material if
there is not a clump immediately near the star, or a similar effect
occurs if there is a clump at small radii but only one one side of the
star.  The effect of making such a model smoother is in fact to fill
in that inner region with denser material, which can increase the mass
of warm dust.  In the smoother models, however, much of the radiation
from such dust (at wavelengths shortward of a few microns) is
extincted and reprocessed back in the far-IR part of the SED (see the
rise in the peak at $\sim$100$\mu$m in Figure \ref{plotdensrat}).  The
result is commonly the relative lowering of the 20$\mu$m and NIR
fluxes as shown in the Figure.
Although there are large variations with viewing angle, the
smoothness of the cocoon has a large enough effect that it can be
observationally probed.  To further illustrate that, we ran many
models with different smooth-to-clumpy ratios and different random
clump distributions. Figure \ref{denshist} shows the results.
Particularly if one can observe a {\it sample} of similar objects, the
mean color of the sample will reflect the smoothness of the
circumstellar dust distribution -- we will demonstrate in the next
section.

\section{Comparison with data}
\label{wccolors}

\begin{figure}
\rplotone{0.5}{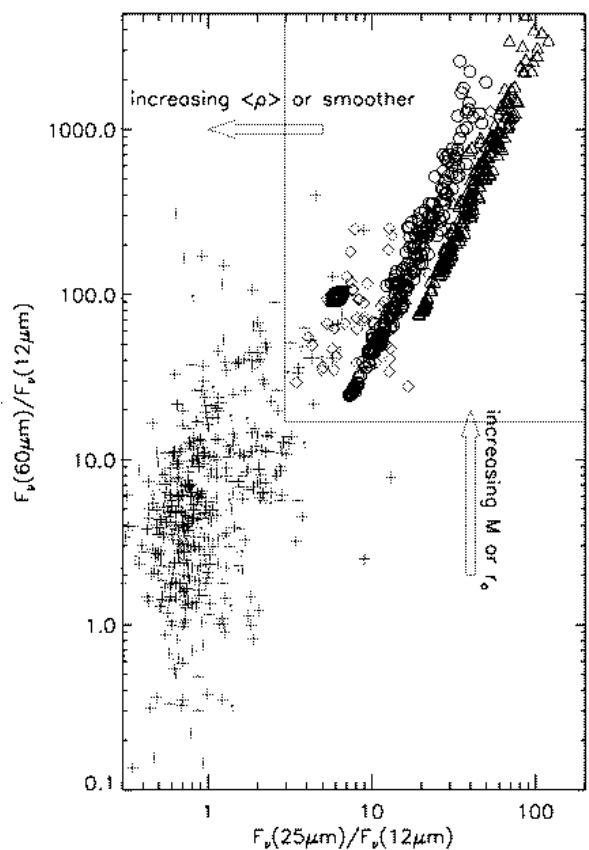}
\caption{\small\label{wcfig} IRAS colors of models and real objects in the
Galaxy.  This color-color diagram is similar to that in WC89, and
shows a random sample of field objects (plus signs) along with
UH\ion{H}{2} regions from \citet{kurtz} and WC89 (diamonds).  Colors
of a model viewed from different angles are shown for smooth-to-clumpy
ratio of 0.03 (triangles), 0.1 (circles), and 0.9 (squares).  Our
models have the WC89 colors of \uchii\ models, but so do smooth models
-- the IRAS bandpasses are not extremely sensitive to dust geometry,
although there are clear trends in this color space with the
smooth-to-clumpy ratio.}
\end{figure}

Now we apply the lessons of the previous section to real datasets, to
determine what we can learn about the physical properties of
ultracompact \ion{H}{2} regions (embedded massive protostars) using our
clumpy models.
\citet{wc89} (hereafter WC89) examined the IRAS colors of ultracompact
\ion{H}{2} regions and determined that objects in a sufficiently red
part of color space were likely to be massive protostars.  This
criterion has been used widely as the starting point for surveys of
protostars and protostellar candidates, including thermal and maser
molecular line targeted observations and imaging from radio continuum
through the infrared.  Although higher resolution more sensitive
infrared data (MSX and {\it Spitzer}/GLIMPSE) is beginning become
available, IRAS is still a cornerstone dataset for studying dusty
Galactic objects, so we calculate and describe the IRAS fluxes and
colors of our models here.  Figure~\ref{wcfig} shows a color-color
diagram similar to that of WC89.  The WC89 color box is plotted along
with the colors of field objects and \uchii\ regions.  The latter are
from the survey of \citet{kurtz} that includes the WC89 sample as a
subset.  Our models have IRAS colors that fit the WC89 criteria, but
smooth models do also.  The colors are shown along different
sightlines for a model with smooth-to-clumpy ratio of 0.03, 0.1, and
0.9.  In this part of parameter space (hot protostars with a
circumstellar dust cocoon about a parsec in size with mean density a
few times 10$^4$~cm$^{-3}$), making the cocoon smoother or increasing
its density while maintaining the same size both result in bluer
[12]-[25] colors, but not much change in [60]-[12].  Decreasing the
total mass without changing the average density (a smaller envelope)
results in bluer [60]-[12] without a large change in [25]-[12].

\begin{figure}
\rplotone{1.0}{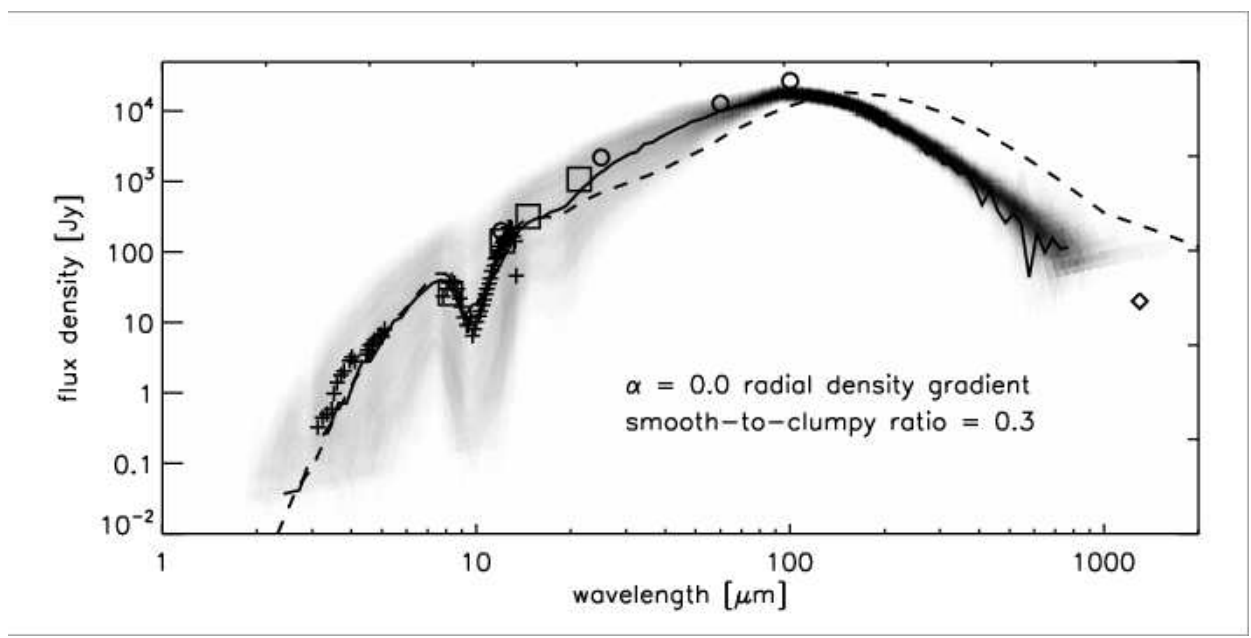}
\caption{\small\label{g5.89a} Comparison of clumpy models to \uchii\ region
G5.89-0.39.  Observations are spectra from F98 (plus signs) and
photometry from IRAS (circles), MSX (squares), and \citet{chini86}
(diamond).  The best-fit sightline of a clumpy model is a solid line
and the range of SEDs for different sightlines of that same clumpy
model (gray scale, darker reflecting a higher density of similar
SEDs).  Also shown is the best-fitting smooth model (dashed line).
The best clumpy sightline fits the data very well; note that we only
tried to fit 1$<\lambda<$15$\mu$m, and that the agreement at longer
wavelengths occurred automatically.  In particular the clumpy model
fits better at 1.3mm datapoint.  The clumpy model is constructed with
a varying aperture size appropriate to the instruments being shown.
This mostly affects the longest and shortest wavelengths.  }
\end{figure}

\begin{figure}
\rplotone{0.8}{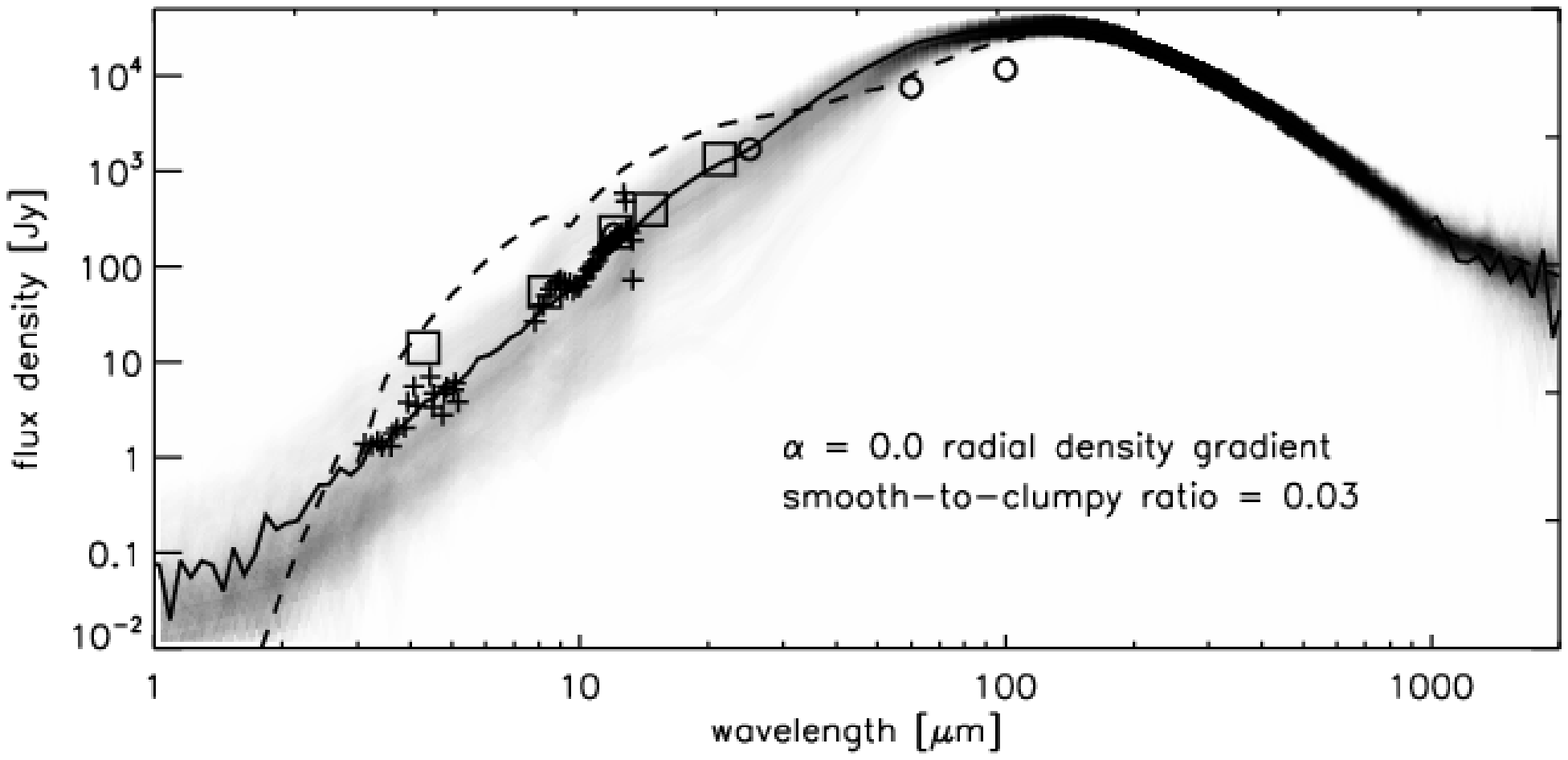}
\rplotone{0.8}{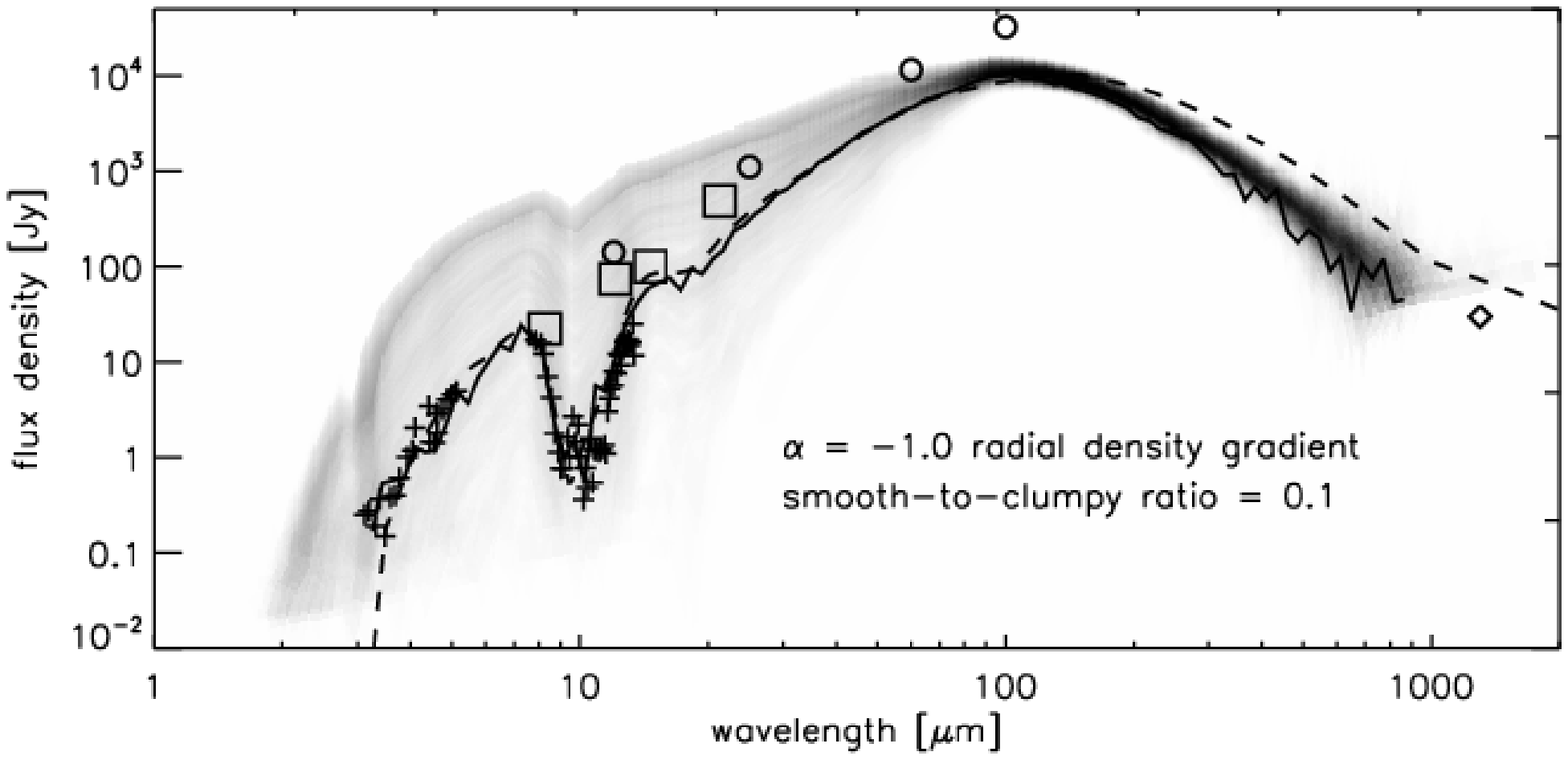}
\rplotone{0.8}{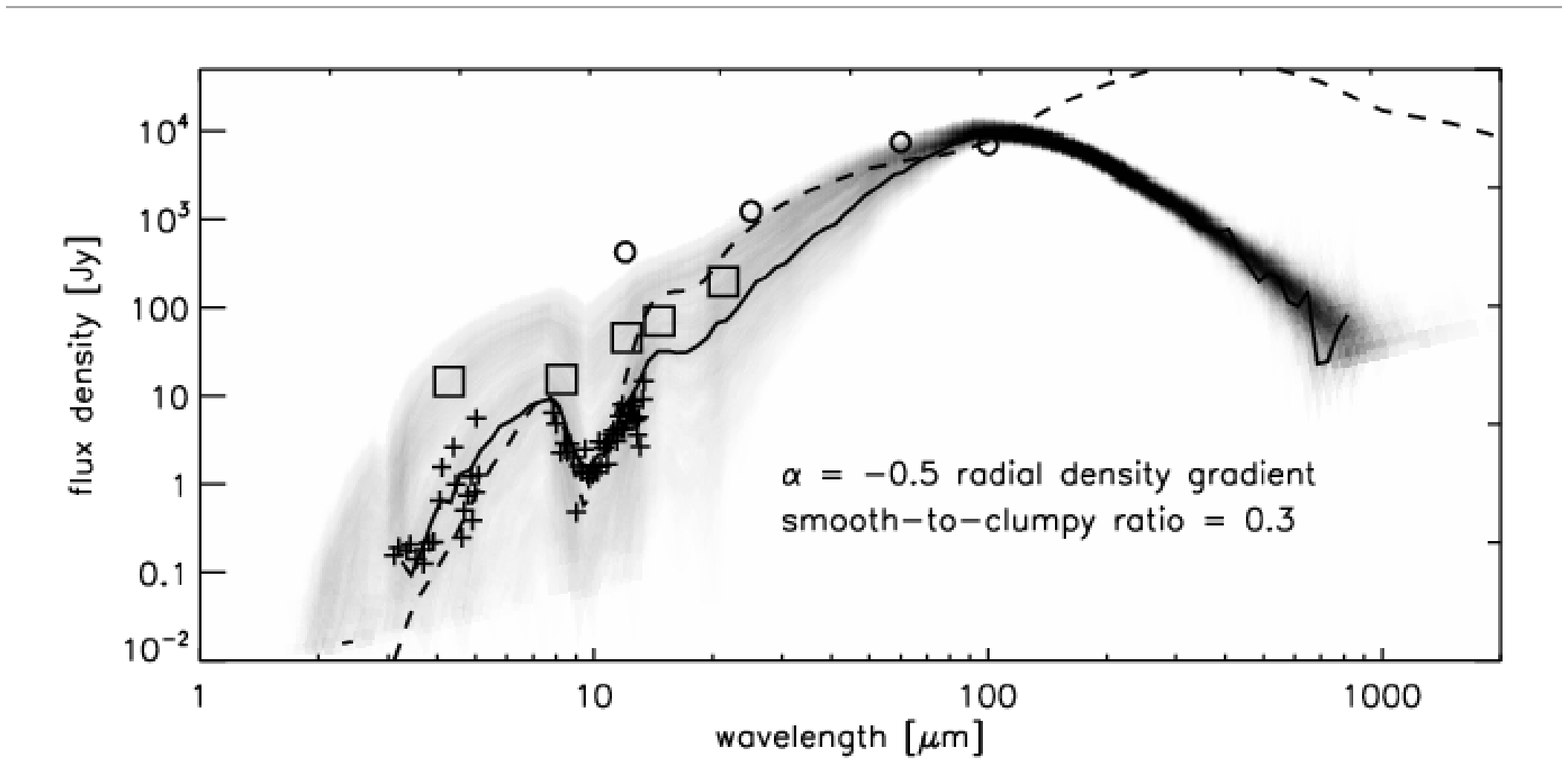}
\caption{\small\label{otherseds} Comparison of clumpy models to \uchii\
regions G29.96-0.02, G34.26+0.15, and G75.78+0.34, annotations as
Fig.~\ref{g5.89a}.}
\end{figure}

Spectroscopy can almost always constrain models more tightly than
photometry, so we next compare against ground-based MIR spectroscopy
of \uchii\ regions by F98.  We begin with G5.89-0.39, an \uchii\ which
was given special attention by those authors and which had been
modeled previously.  Figure~\ref{g5.89a} shows data from F98, IRAS,
and MSX along with a smooth model and our canonical clumpy model.  We
compared the data with our grid of smooth models and confirm that the
parameters similar to those given by F98 (they used $r_{in}$=0.032pc,
$r_{out}$=2.6pc, $\alpha$=0.0, $M$=70000$M_\sun$) produce the best
fit, but that fit is still not perfect, in particular shortward of
8$\mu$m and around 20$\mu$m.  We then ran a large number ($\sim$600)
of clumpy models with different clump distributions, inner radii,
average radial density gradients, and smooth-to-clumpy ratios (but
fixed outer radius), and found the best-fitting sightline by
least-squared difference between the data and model.  {\bf We fit only
the F98 data between 3 and 13$\mu$m, but the best-fitting models also
fit the longer wavelength data, with no further adjustment.}  In
particular, the slope of the SED around 20$\mu$m and the behavior
shortward of 8$\mu$m resemble the data much more closely than the
smooth model.  The best-fitting smooth and clumpy models' parameters
are listed in Table~\ref{datatable}.  The figure also shows the SEDs
for other sightlines of the same clumpy model as grayscale, showing
that if this is indeed an accurate model of G5.89, that it would look
rather different from a different vantage point.
Figure~\ref{otherseds} shows three other objects from F98 with
different types of SEDs (NIR intensity and silicate emission or
absorption), all of which are fit better by clumpy models than smooth
ones (see $\chi^2$ values and models parameters in
Table~\ref{datatable}).  In some cases, the model flux falls short of
the observations at longer wavelengths, in particular at 100$\mu$m.
At the size of the 100$\mu$m IRAS beam, the Galaxy is very confused,
and nearby sources almost certainly contribute some heating of the
outer dust cocoon, increasing the far-IR flux.

\begin{deluxetable}{llrlccccc}
\rotate
\tablecolumns{9}
\tablewidth{0cm}
\tabletypesize{\small}
\tablecaption{\small\label{datatable} Parameters of 1-D and 3-D models fitting data}
\tablehead{\colhead{object} & \colhead{model} & \colhead{density} & 
\colhead{r$_i$}  & \colhead{r$_o$} & \colhead{smooth-to} & 
\colhead{mass} & \colhead{Average ${\rm A_V}$} & \colhead{$\chi^2$} \\
& & \colhead{index $\alpha$} & \colhead{[pc]} 
& \colhead{[pc]} & \colhead{clumpy} & \colhead{M$_\sun$}
& \colhead{to star} & }
\startdata
G5.89-0.39  &smooth unconstrained              &+1.5 & 0.00039 & 0.39 &1    &1700              & 109 & 2.7 \\
            &smooth $\alpha\leq0,r_{out}\leq7$ & 0.0 & 0.0059  & 5.9  &1    &2.3$\times$10$^5$ & 109 & 6.1 \\
            &clumpy                            & 0.0 & 0.0001  & 2.5  &0.3  &50000             & 194 & 2.5 \\
G29.96-0.02 &smooth unconstrained              &+0.5 & 0.0039  & 1.1  &1    &3500              &  50 & 5.8 \\
            &smooth $\alpha\leq0,r_{out}\leq7$ & 0.0 & 0.0039  & 3.9  &1    &51000             &  50 & 7.1 \\
            &clumpy                            & 0.0 & 0.0001  & 2.5  &0.03 &50000             & 178 & 1.1 \\
G34.26+0.15 &smooth unconstrained              &-0.5 & 0.0039  & 3.8  &1    &1.3$\times$10$^5$ & 150 & 11. \\
            &clumpy                            &-1.0 & 0.003   & 2.5  &0.1  &50000             & 259 & 2.5 \\
G75.78+0.34 &smooth unconstrained              &+0.5 & 0.0019  & 19.  &1    &2.1$\times$10$^7$ & 100 & 4.7 \\
            &smooth $\alpha\leq0,r_{out}\leq7$ & 0.0 & 0.0039  & 3.9  &1    &2.0$\times$10$^5$ & 200 & 16. \\
            &clumpy                            &-0.5 & 0.0001  & 2.5  &0.3  &50000             & 238 & 1.7 \\
\enddata                                     
\end{deluxetable}

\begin{figure}
\begin{center}
\includegraphics*[width=2.3in]{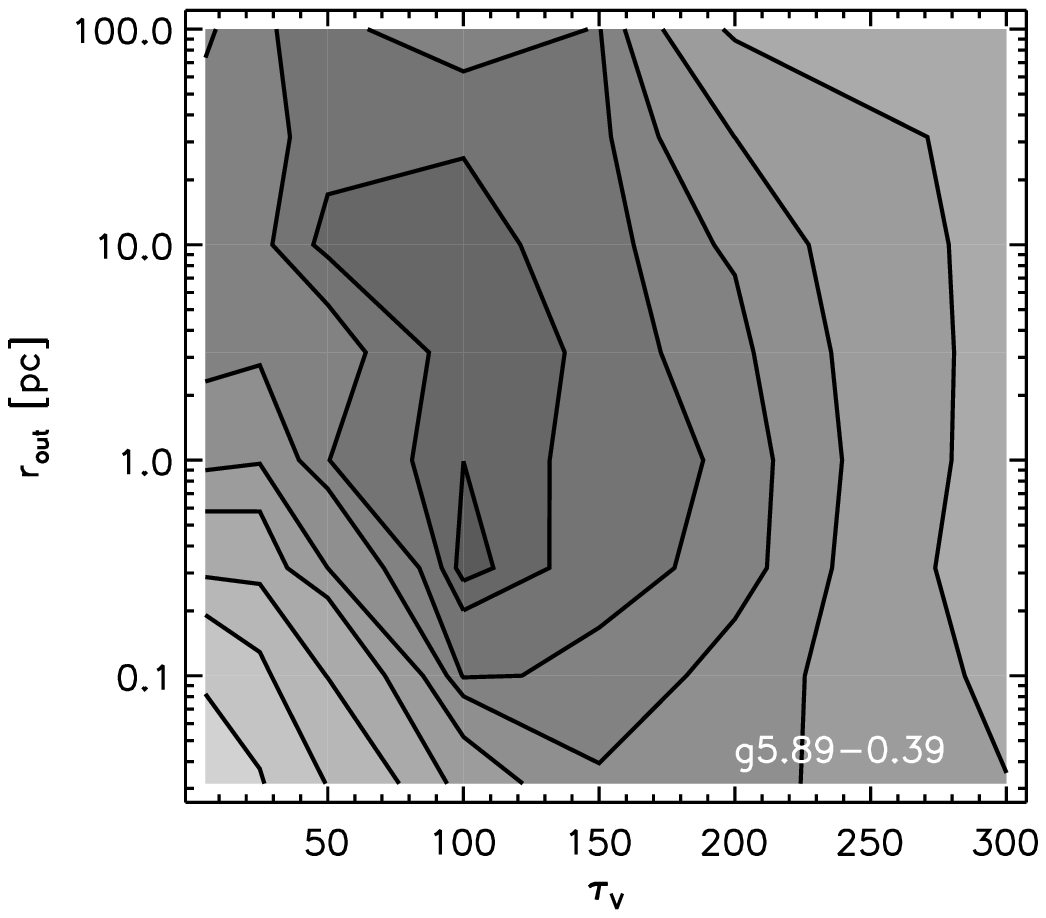}
\includegraphics*[width=2.3in]{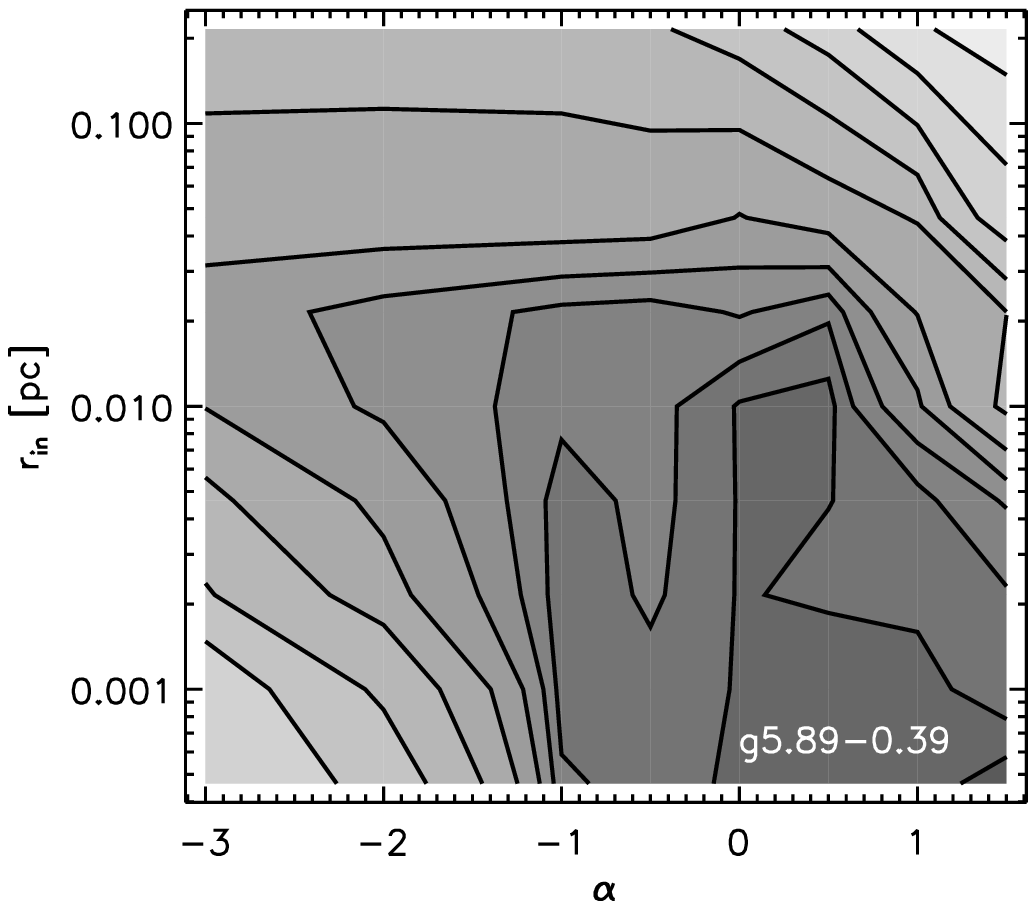}\\
\includegraphics*[width=2.3in]{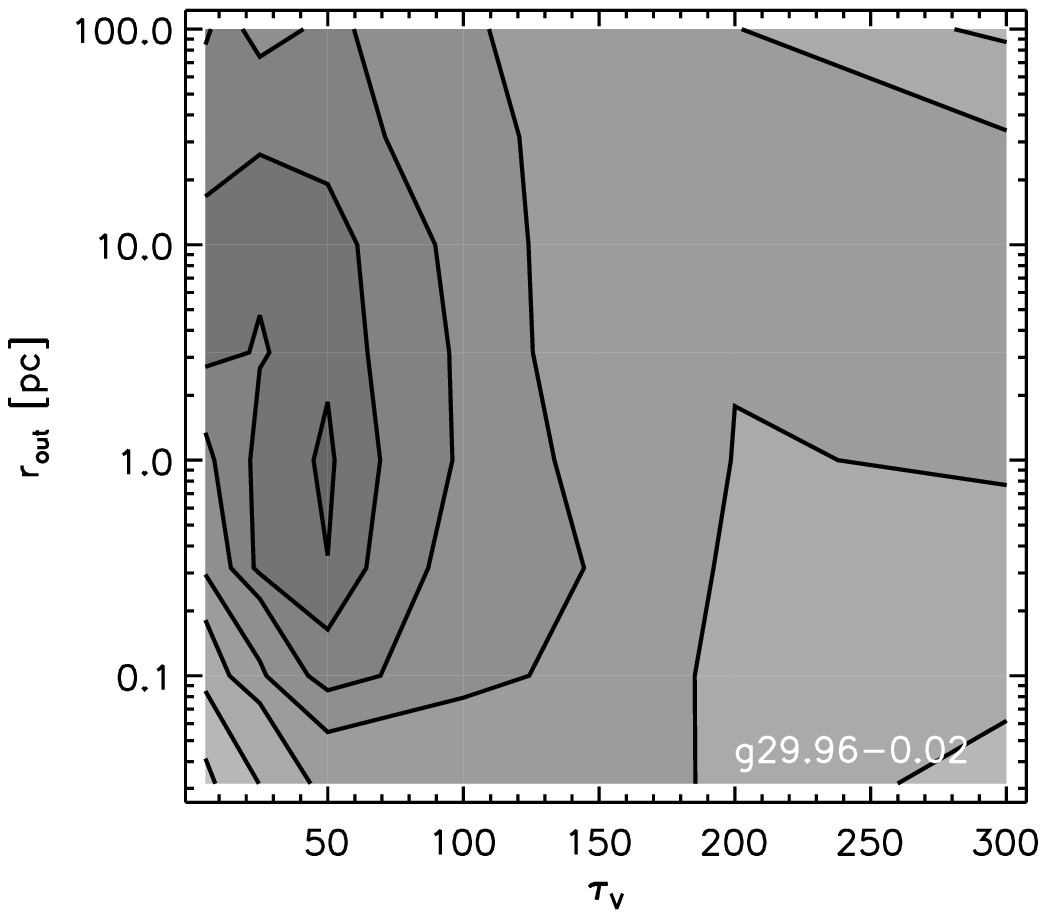}
\includegraphics*[width=2.3in]{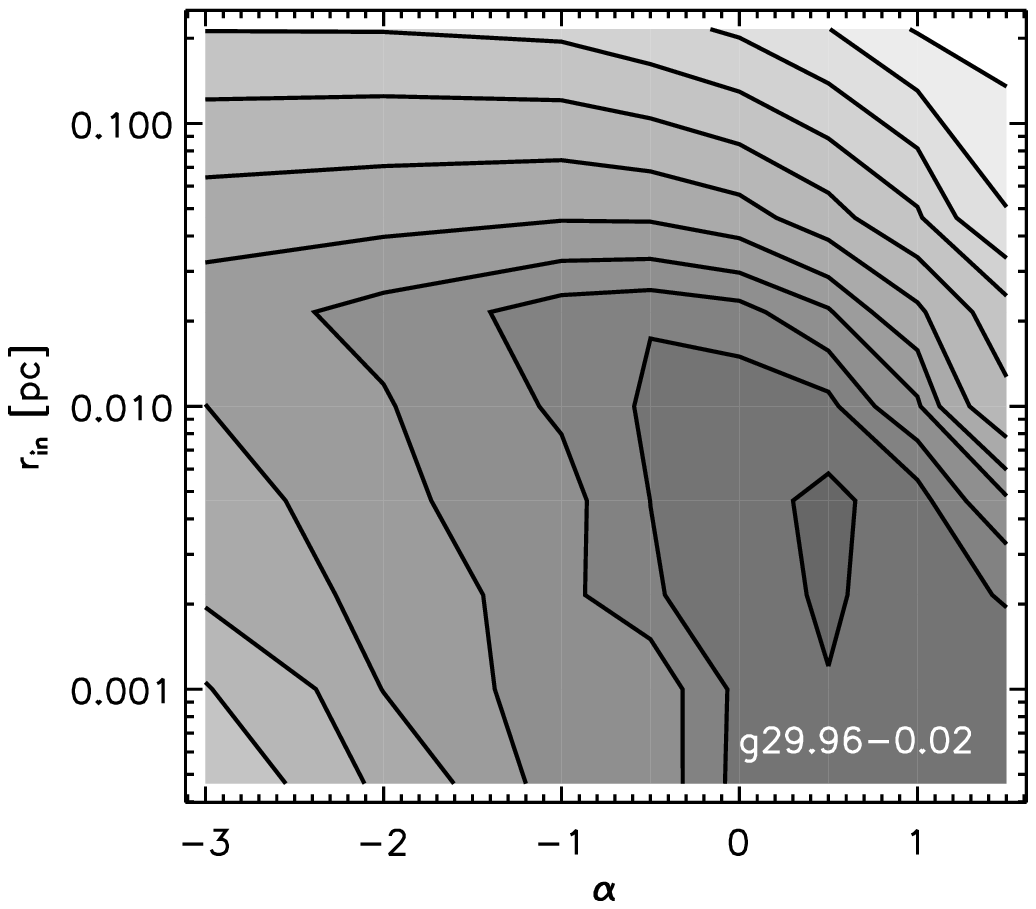}\\
\includegraphics*[width=2.3in]{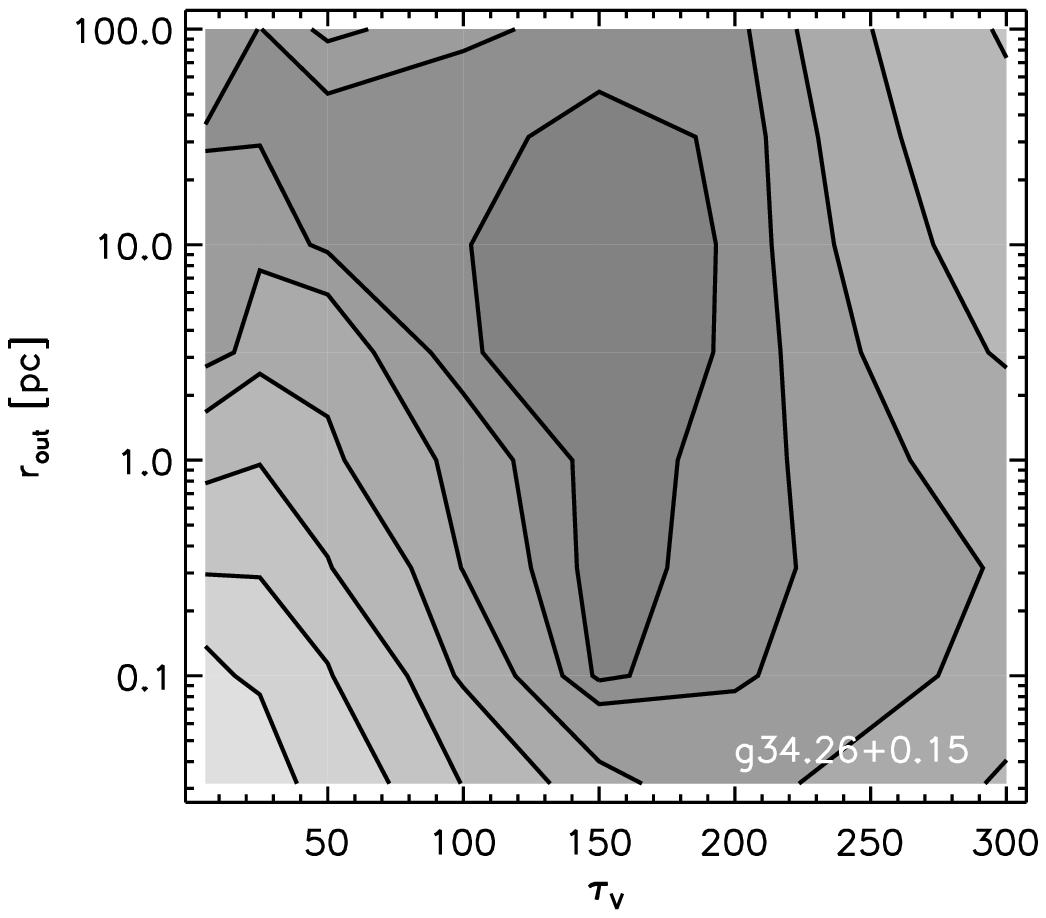}
\includegraphics*[width=2.3in]{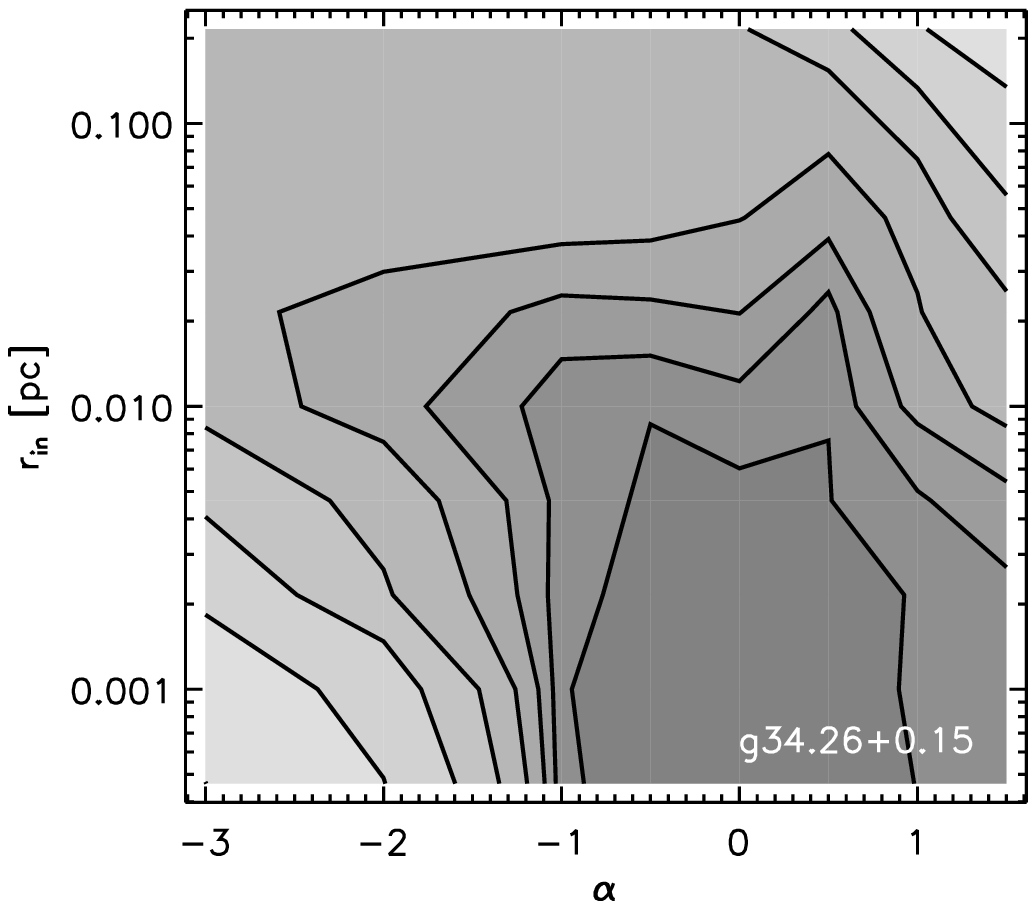}\\
\includegraphics*[width=2.3in]{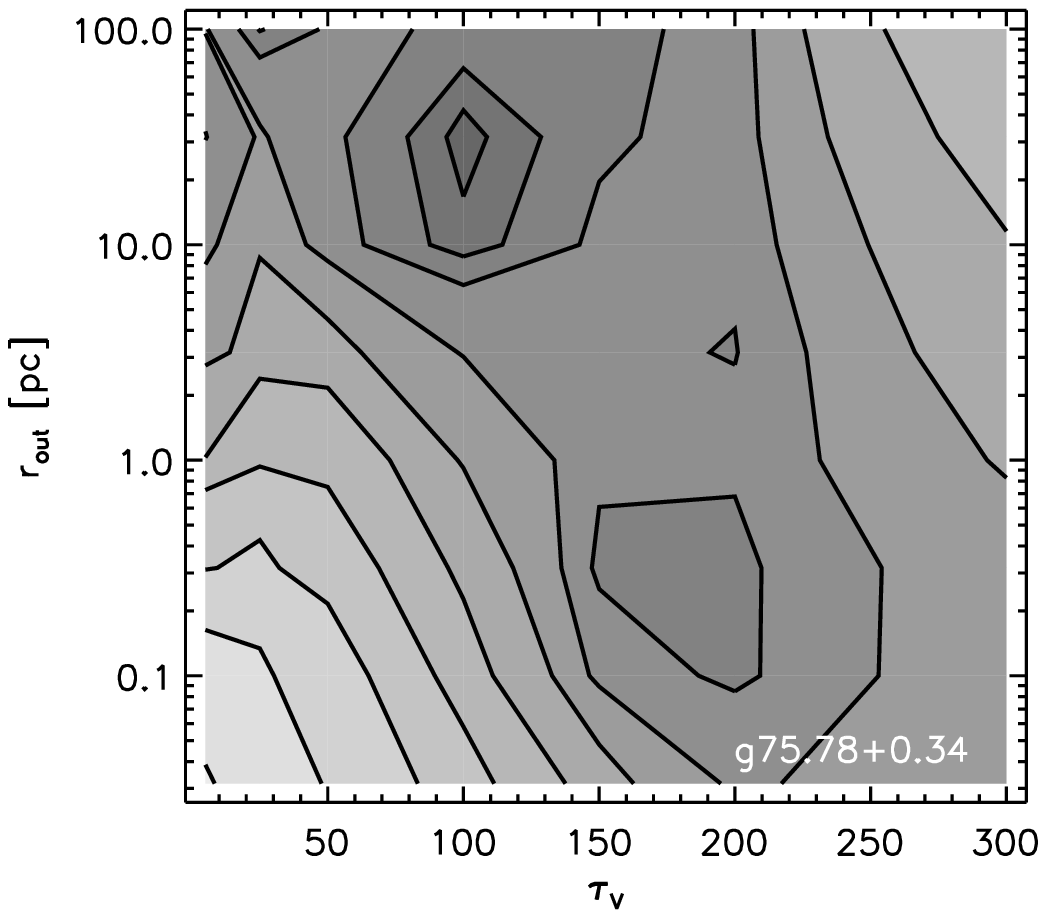}
\includegraphics*[width=2.3in]{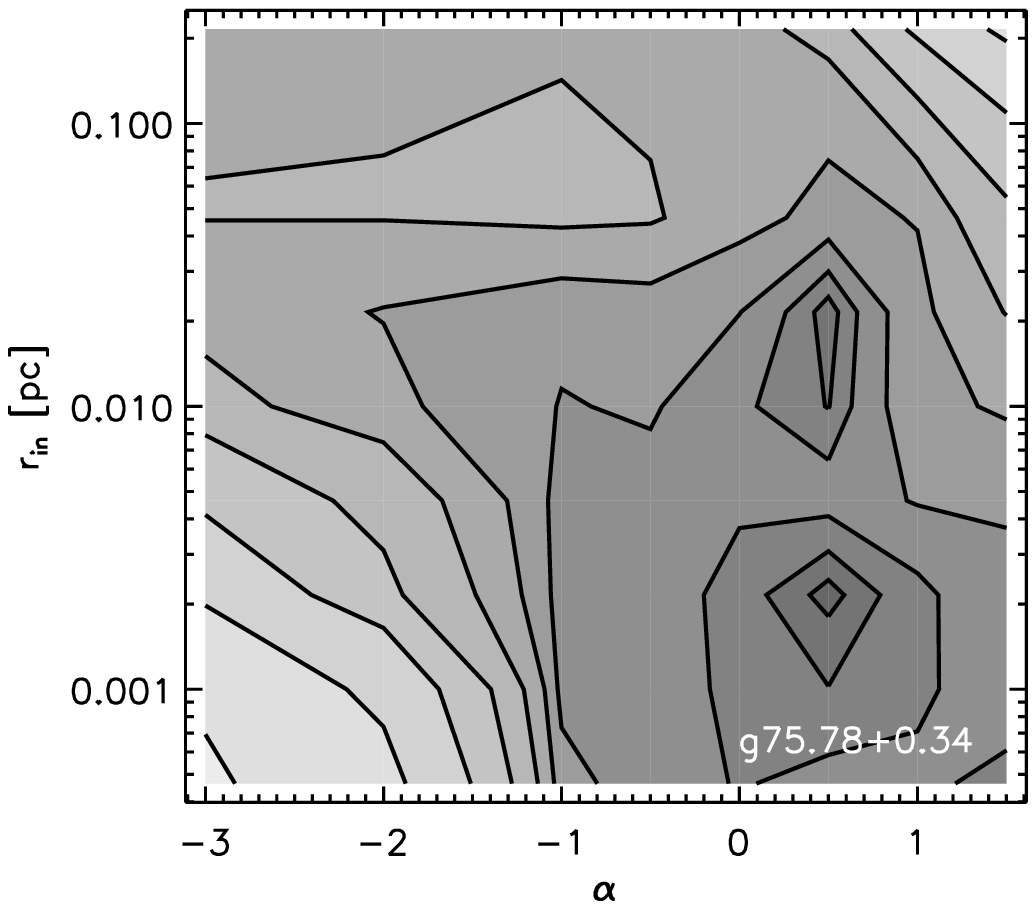}\\
\end{center}
\caption{\small\label{datadusty} Contours of goodness-of-fit of 1-D smooth models 
to four \uchii{} regions. The models favor large outer radii and often often 
large inner radii as well -- see text for discussion.}
\end{figure}

\begin{figure}
\begin{center}
\includegraphics*[width=2.3in]{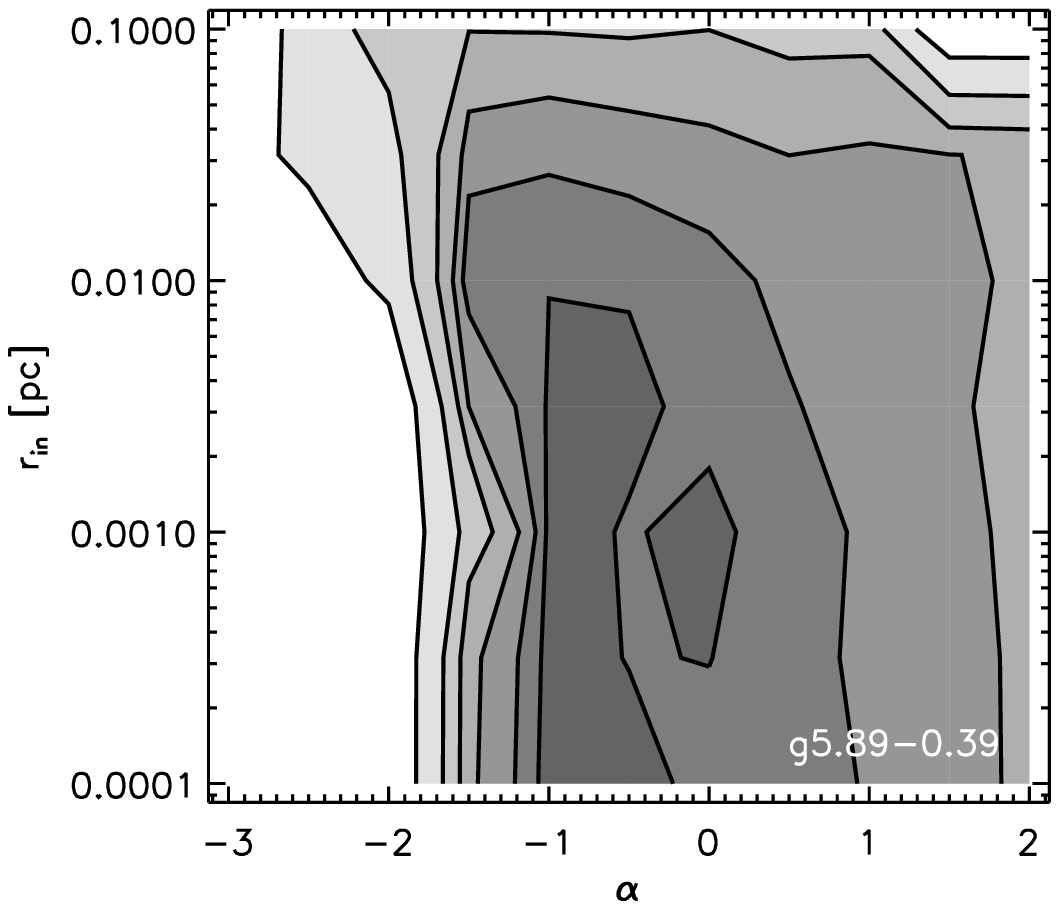}
\includegraphics*[width=2.3in]{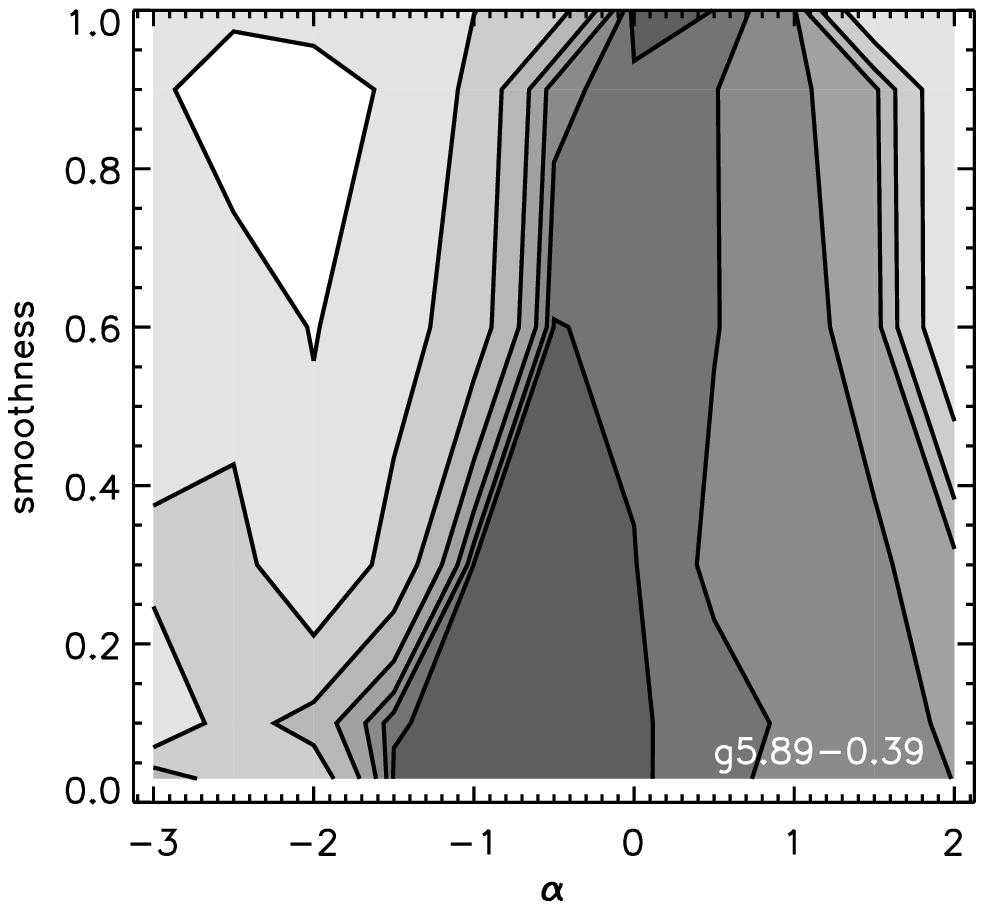}
\\			  
\includegraphics*[width=2.3in]{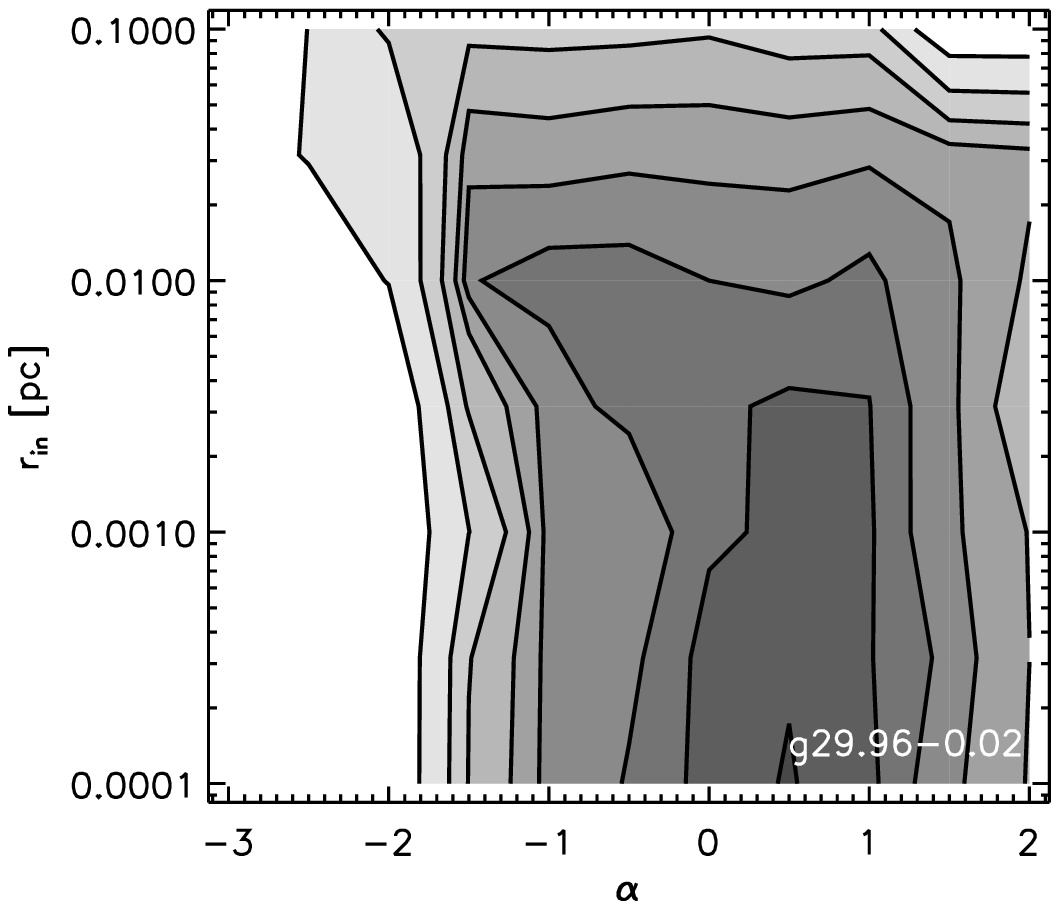}
\includegraphics*[width=2.3in]{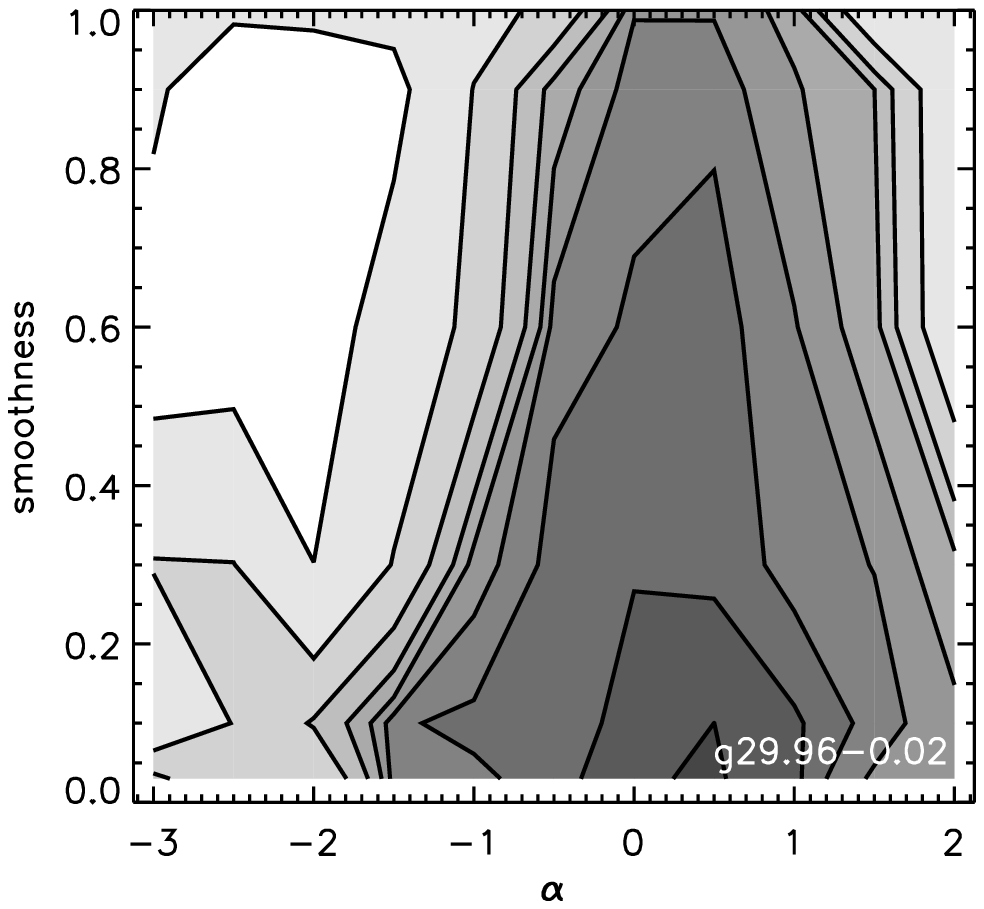}
\\			  
\includegraphics*[width=2.3in]{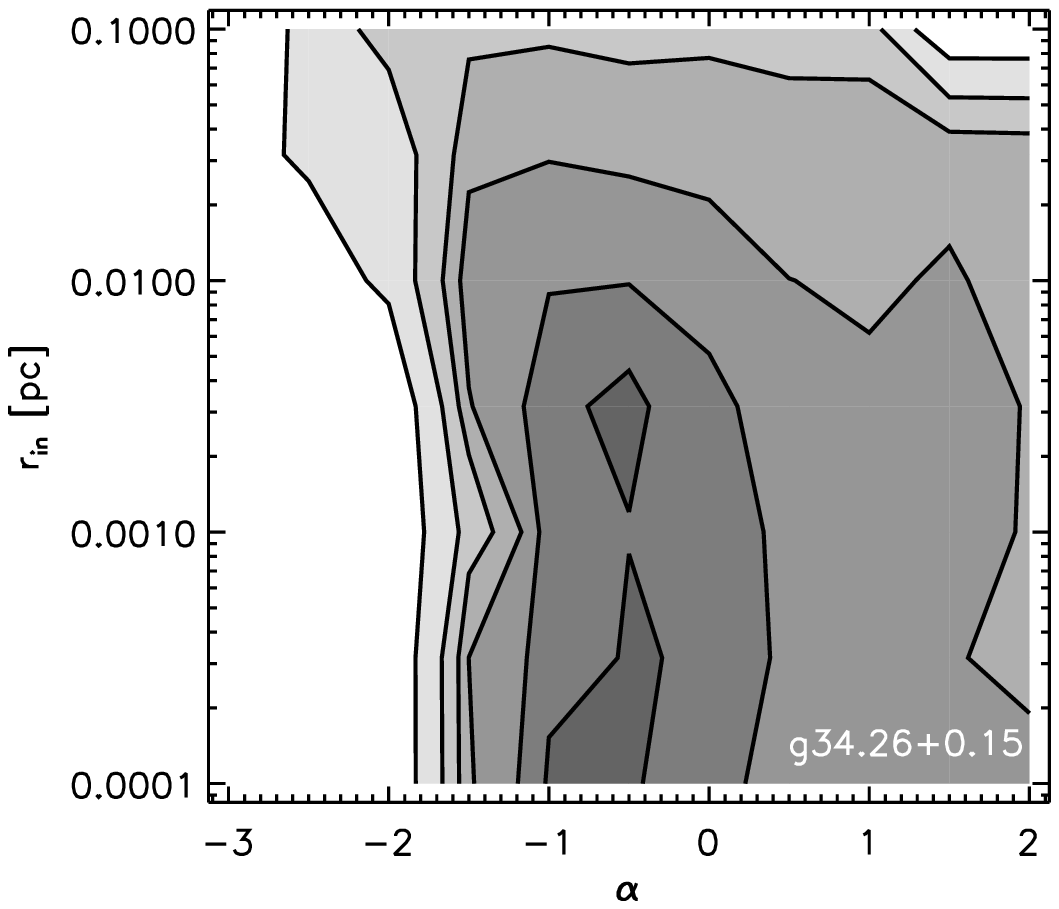}
\includegraphics*[width=2.3in]{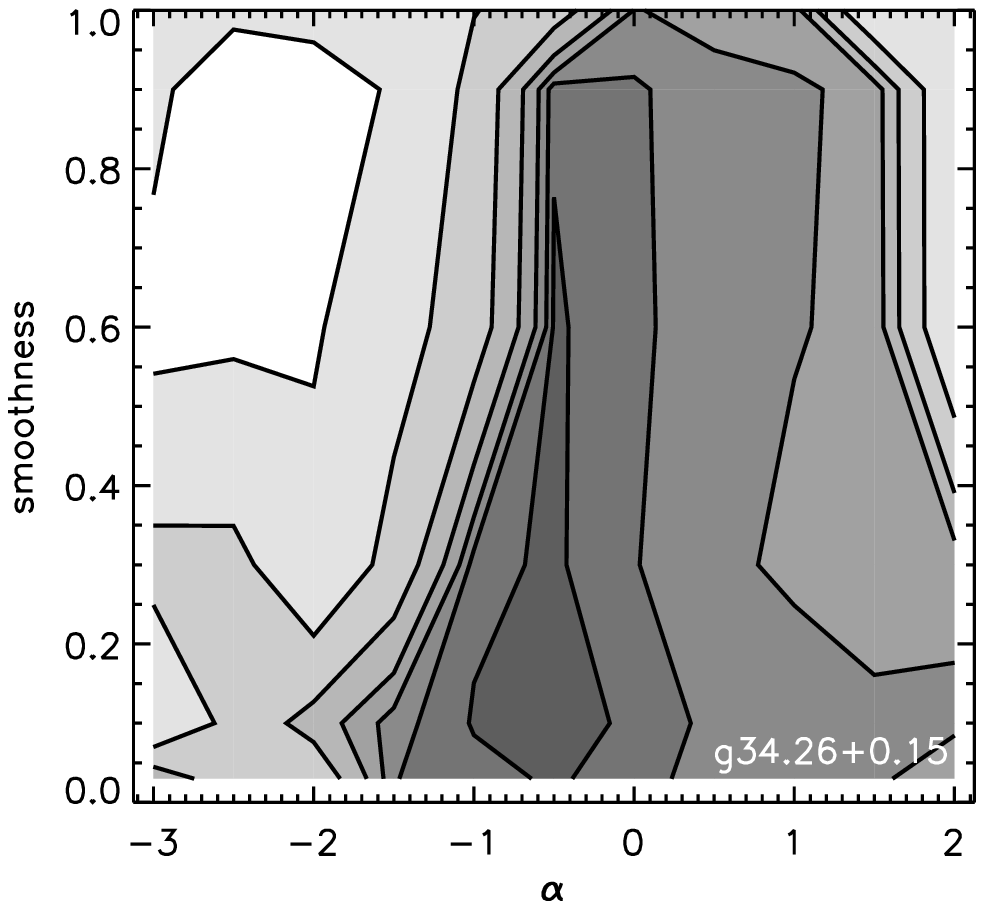}
\\			  
\includegraphics*[width=2.3in]{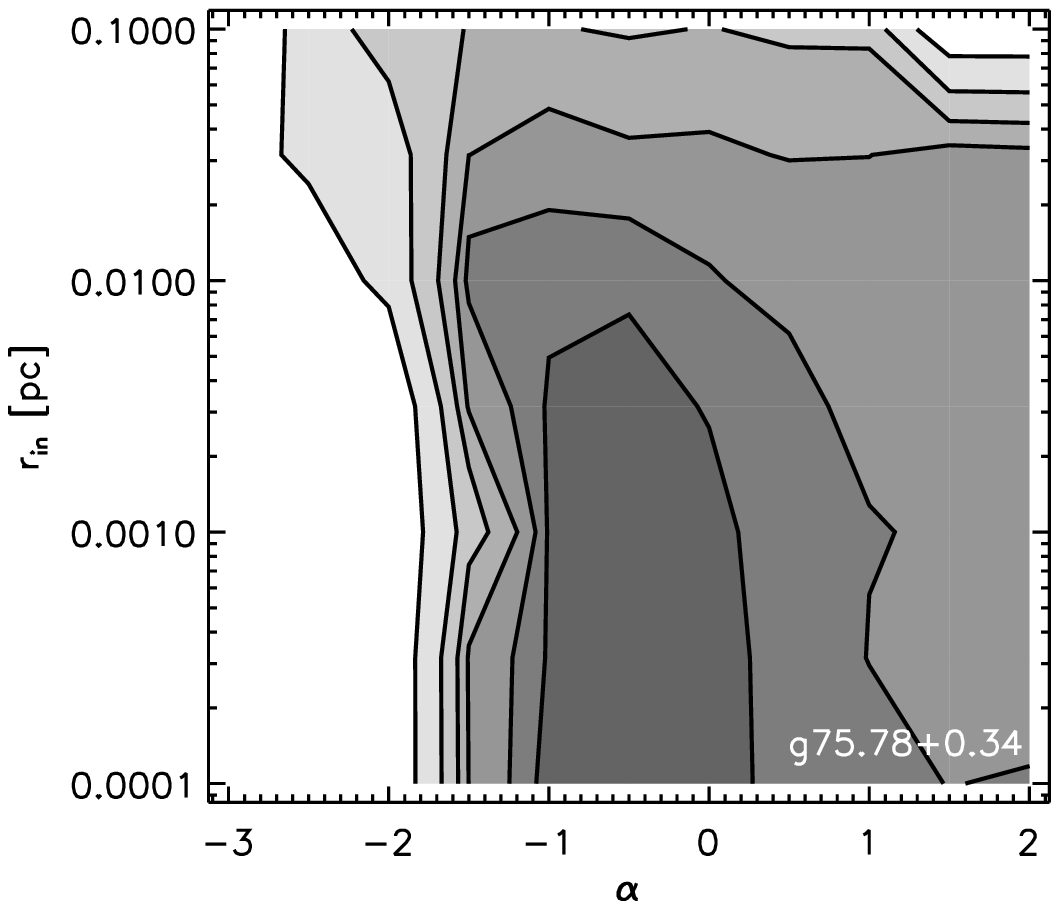}
\includegraphics*[width=2.3in]{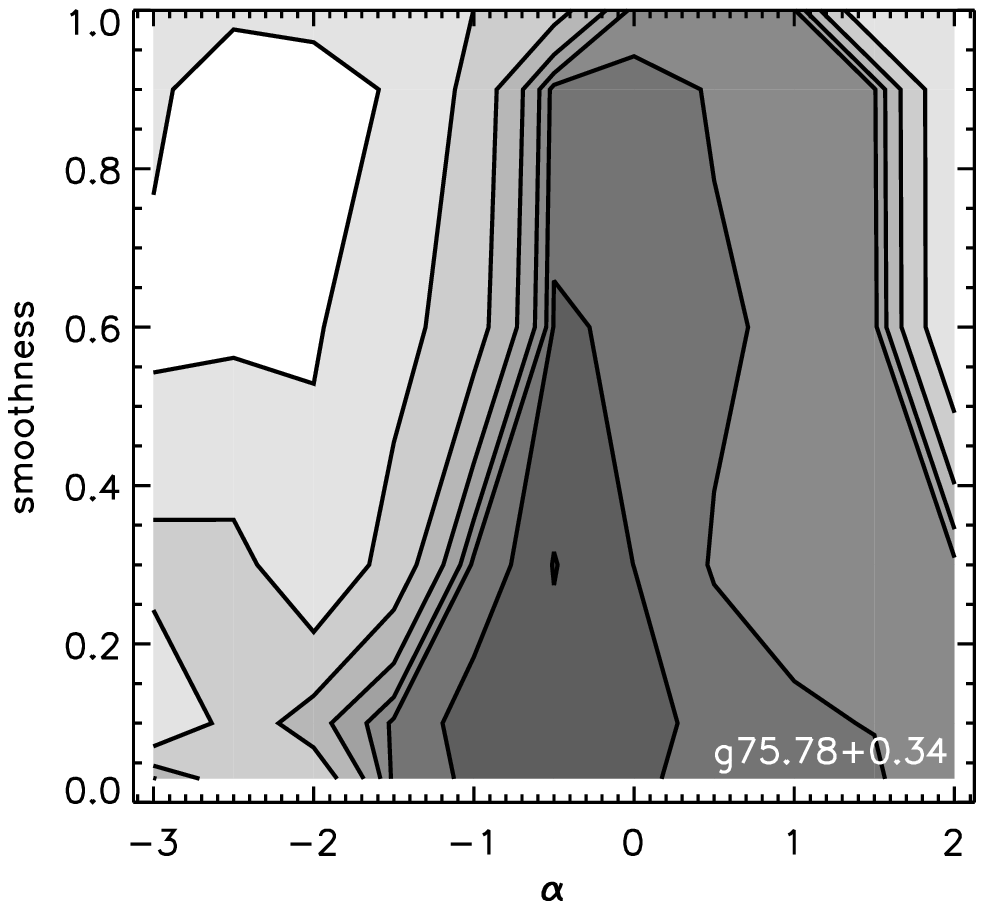}
\end{center}
\caption{\small\label{dataclumpy} Contours of goodness-of-fit of 3-D clumpy
models to four \uchii{} regions. The models favor relatively flat
radial profiles and low smooth-to-clumpy ratios, and best fits tend to
have lower $\chi^2$ than the best fitting 1-D models -- see text for
discussion.}
\end{figure}

Figures~\ref{datadusty} and \ref{dataclumpy} show contours of
goodness-of-fit for the smooth and clumpy models, respectively.
Smooth models have the same difficulties fitting real data that 
they did in fitting SEDs from 3-D models above:  The better-fitting
models have flat density profiles or even ones that increase 
as a function of radius.  The outer radii are uncomfortably large, 
and even the best-fitting models have $\chi^2\gtrsim 7$.  

Clumpy models with a smooth-to-clumpy ratio of $\lesssim$50\% are
preferred, with a loose preference for the range 0.1--0.3.  Models
with highly evacuated voids (3\% smooth) have more difficulty
reproducing the 5$\mu$m flux and short-wavelength shape of the SED
seen in some objects.  On the other hand, models too smooth begin to
show the shortcomings of the completely smooth models and also fit
less well.  When considering the radiation intensity in the interior
of an externally lit cloud, \citet{bethell} found that a
smooth-to-clumpy ratio of 0.333 in a hierarchically clumped model
produced a similar internal intensity to an externally lit turbulent
hydrodynamic simulation -- corroborating evidence of a low smooth
fraction.

Finally, we note that models with fairly flat angle-averaged radial
density profiles often fit better than those with a strong radial
gradient, but that the clumpy models with negative gradients
(decreasing density with radius) fit better than smooth models with
negative gradients.  Both findings are important: some \uchii{}
regions such as G5.89 are well-described by a star in a clumpy
molecular cloud, without any particularly concentrated envelope.  This
is consistent with the notion that \uchii{}s are relatively evolved
protostars, perhaps with accretion already halted \citep{edaraa}.
\footnote{ We note that G5.89-0.39 is a multiple source
\citep{feldt03,sollins04}, which is consistent with our interpretation as
an illuminating source embedded in a molecular cloud without any
particular radial density gradient.  We are currently investigating
models with multiple sources, and although beyond the scope of this
paper, the SEDs of small clusters show many similar effects to those
discussed here. }  
However, radial density profiles have been invoked
to explain the submm spatial intensity distribution of high-mass
protostellar objects \citep[e.g.][]{hatchell}.  
A full analysis of submm radial intensity
profiles is beyond the scope of this paper, but preliminary
investigation finds that clumpy models with $\alpha$ between -0.5 and
-1.0 fit the ``non-peaked'' SCUBA sources of \citep{hatchell} as well
as 1-D models, and clumpy models with $\alpha\simeq$-1.5 fit the
``peaked'' sources very well, without needing to invoke extra emission
in a ``core'' as was required by those authors with 1-D models.

\begin{figure}[h]
\rplotone{1.}{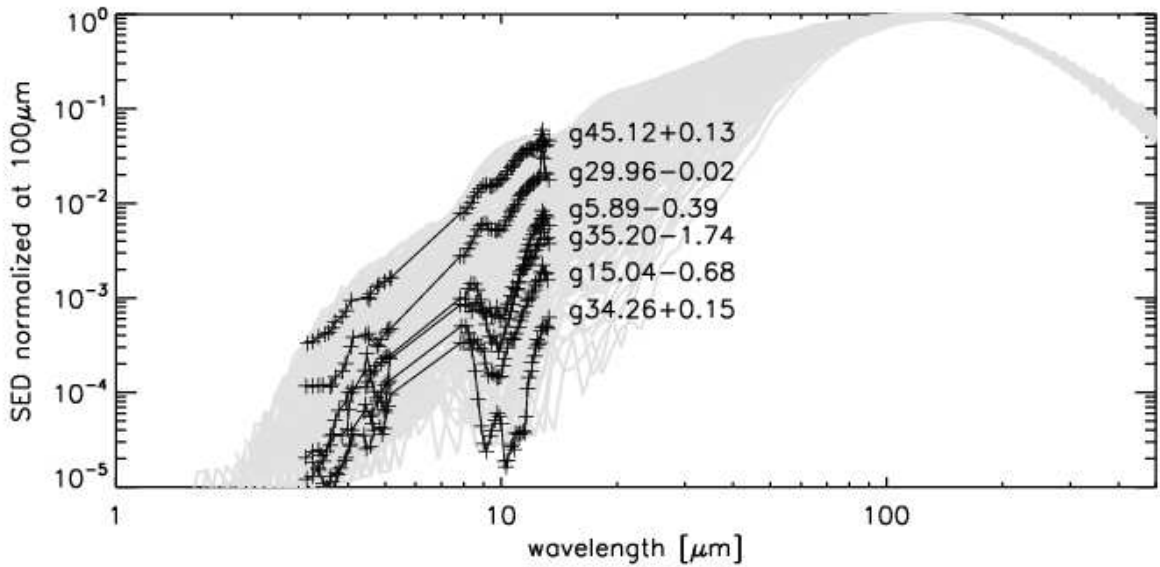}
\caption{\small\label{allmodel} The same clumpy model that fits G5.89
(Fig.~\ref{g5.89a}, top panel) can in fact fit F98's entire sample of
\uchii\ regions reasonably well!  Data for several regions are
here plotted as connected plus signs on top of the gray-scale range
of sightlines for a single clumpy model.}
\end{figure}

In Figure~\ref{allmodel}, we show something even more remarkable: the
single clumpy model that best fits G5.89 can in fact do a reasonable
job fitting F98's entire collection of \uchii\ regions!  We normalized
each region's data to the 100$\mu$m point, mostly to correct for
distance, but this also corrects for the difference in spectral type
between the regions.  For a deeply embedded object, the shape of the
output (observable) SED shows very little variation whether the cocoon
is heated by an O7 star or a B2 star, because the NIR/MIR are
dominated by reprocessed emission in both cases \citep[see discussion
in][]{pIII}.  The total luminosity of the object naturally scales with
the bolometric luminosity of the illuminating star.  It is very
interesting that all of these young protostars are consistent with the
same model seen from different directions.  Observational tests must
be done using ensembles of objects, and some care should be taken not to
overinterpret differences between individual objects as differences in
evolutionary state or cocoon mass.

\section{Conclusions}
\label{conclusions}

We use 3-D radiative transfer models to examine the importance of an
inhomogeneous circumstellar envelope on the observed properties of
young embedded massive stars.  The spectral energy distributions of
3-D models vary significantly with viewing angle, and differ
significantly from 1-D models.  Some of our most important findings are
as follows:
\begin{list}{$\bullet$}{}
\item{The SEDs of embedded stars and young stellar objects are
significantly affected by clumping of the dust at wavelengths
$\lesssim$100$\mu$m.  Longer-wavelength emission is primarily
determined by the total mass of thermally emitting dust rather than
its distribution, and therefore less affected by clumping.}
\item{Silicate features (the strongest and most noticeable at 10$\mu$m)
can be seen in emission or absorption in {\it same object}, depending on 
viewing angle.}
\item{Attempts to fit 3-D clumpy objects with 1-D models can
mis-estimate the size of the region and total mass by more than an
order of magnitude.  The derived values of the region size, mass,
central object luminosity, and optical depth are very often incorrect
by a factor of 2--3.}
\item{The typical or average SED (considering all viewing angles) is
sensitive to the overall smoothness of the dust distribution, or the
smooth-to-clumpy ratio.  The {\uchii}s observed by F98 are
best fit by models that are $\lesssim$50\% smooth. }
\item{Our 3-D models do a better job at fitting the entire SED of
{\uchii}s than previous 1-D models, and can easily explain the
variation in \uchii\ SEDs as differences in viewing angle of a clumpy
medium.}
\item{Using 3-D clumpy models tends to allow the data to be fit 
well with more peaked density profiles than the best 1-D fits, 
which may agree better with SCUBA imaging data.  Some \uchii{}
are best fit with flat density profiles, whether one uses 
1-D or 3-D models -- these may be more evolved objects 
best understood as just a hot star embedded in a non-accreting cloud.}
\end{list}

It is becoming clear that models with more than one dimension are
necessary to understand the observed properties of interstellar and
circumstellar dust and gas.  We have shown here that clumpiness can
severely affect the emergent spectral energy distribution of embedded
massive stars and \uchii\ regions.  Previous work by us and others has
shown that at least two-dimensional models are critical to
understanding accreting protostars; clumpiness in such YSOs may have
further effects.  In the general ISM, clumpiness affects the radial
profiles and line ratios observed in \ion{H}{2} regions
\citet{wood05}.  Clumping has been invoked to explain the observed
silicate feature in AGN torii \citet{agn}.  Models such as our
code are invaluable to determine which physical
parameters are observationally accessible in complex and multi-phase
systems.

\vspace{2cm}
\centerline{\bfseries Acknowledgements}

During this work, RI was supported by the GLIMPSE
{\it Spitzer} Legacy program (E Churchwell, P.I.), BW by NASA LTSA
(NAG5-8933), and KJ an NSF AST postdoctoral fellowship.  This
publication used data from 2MASS, MSX, and IRAS, and made extensive
use of NASA's ADS abstract service. The Two Micron All Sky Survey is a
joint project of the University of Massachusetts and IPAC/Caltech,
funded by NASA and NSF.  Processing of Midcourse Space Experiment Data
was funded by the Ballistic Missile Defense Organization with
additional support from NASA.  The data are served byt the NASA/IPA
Infrared Science Archive operated by JPL/Caltech.

We thank Ed Churchwell, John Mathis, and Jon Bjorkman for useful
discussions on massive star formation, interstellar clumps, and
radiative transfer.

\let\oldthebibliography=\thebibliography
\let\endoldthebibliography=\endthebibliography
\renewenvironment{thebibliography}[1]{
  \begin{oldthebibliography}{#1}
    \setlength{\parskip}{0ex}
    \setlength{\itemsep}{0ex}}
{ \end{oldthebibliography} }

\end{document}